\documentclass{article}

\usepackage{arxiv}
\usepackage[utf8]{inputenc} 
\usepackage[T1]{fontenc}    
\usepackage{hyperref}       
\usepackage{booktabs}       
\usepackage{amsfonts}       
\usepackage{nicefrac}       
\usepackage{microtype}      
\usepackage{lipsum}
\usepackage{graphicx}
\usepackage{float}
\usepackage{wrapfig}
\restylefloat{figure}
\usepackage{multirow}
\usepackage[table,xcdraw]{xcolor}
\usepackage[bottom]{footmisc}
\usepackage{etoolbox}
\usepackage{adjustbox}
\usepackage{booktabs}
\usepackage{array}
\usepackage{hhline}
\usepackage{caption}
\usepackage{subcaption}
\usepackage[normalem]{ulem}
\useunder{\uline}{\ul}{}
\usepackage{longtable}
\usepackage{tabularx}
\usepackage{url}
\newcolumntype{L}[1]{>{\raggedright\let\newline\\\arraybackslash\hspace{0pt}}m{#1}}
\newcolumntype{C}[1]{>{\centering\let\newline\\\arraybackslash\hspace{0pt}}m{#1}}
\newcolumntype{R}[1]{>{\raggedleft\let\newline\\\arraybackslash\hspace{0pt}}m{#1}}

\title{A Survey of Benchmarks to Evaluate Data Analytics for Smart-* Applications}

\author{
    Athanasios~Kiatipis \\
    Fujitsu Technology Solutions GmbH\\
    Munich, Germany\\
    \texttt{a.kiatipis@uniwa.gr} \\
    \And
    Alvaro Brandon \\
    Universidad Politécnica de Madrid\\
    Madrid, Spain\\
    \texttt{abrandon@fi.upm.es} \\
    \And
    Rizkallah Touma \\
    Barcelona Supercomputing Center\\
    Barcelona, Spain\\
    \texttt{rizkallah.touma@yahoo.com} \\
    \And
    Pierre Matri \\
    Universidad Politécnica de Madrid\\
    Madrid, Spain\\
    \texttt{pmatri@fi.upm.es} \\
    \And
    Micha{\l}~Zasadzi{\'n}ski \\
    CA Technologies \\
    Barcelona, Spain \\
    \texttt{michal.zasadzinski@gmail.com} \\
    \And
    Linh Thuy Nguyen \\
    IMT Atlantique, Nantes, France\\
    Inria, Rennes Br. Atl., France \\
    \texttt{thuy-linh.nguyen@inria.fr} \\
      \And
    Adrien Lebre \\
    IMT Atlantique, Nantes, France\\
    Inria, Rennes Br. Atl., France\\
    \texttt{adrien.lebre@inria.fr} \\
    \And
    Alexandru Costan \\
    INSA Rennes, Rennes, France\\
    Inria, Rennes Br. Atl., France\\
    \texttt{alexandru.costan@inria.fr} \\
}

\begin{document}
\maketitle

\markboth{Arxiv}%
{Kiatipis \MakeLowercase{\textit{et al.}}: A Survey of Benchmarks to Evaluate Data Analytics for Smart-* Applications}

\begin{abstract}
The growth of ubiquitous sensor networks at an accelerating pace cuts across many areas of modern day life. They enable measuring, inferring, understanding and acting upon a wide variety of indicators, in fields ranging from agriculture to healthcare or to complex urban environments. The applications devoted to this task are designated as Smart-* Applications. They hide a staggering complexity, relying on multiple layers of data collection, transmission, aggregation, analysis and also storage, both at the network edge and on the cloud. Furthermore, Smart-* Applications raise additional specific challenges, such as the need to process and extract knowledge from diverse data, which is flowing at high velocity in near real-time or in the heavily distributed environment they rely on. How to assess the performance of such a complex stack, when faced with the specifics of \mbox{Smart-*} Applications, remains an open research question. In this article, the key specific characteristics and requirements of Smart-* Applications are initially detailed. Afterwards, for each of these requirements, there is a description of the benchmarks one can use to precisely evaluate the performance of the underlying systems and technologies. Finally, an identification of future research directions related to identified open issues for benchmarking Smart-* Applications is performed.
\end{abstract}

\keywords{IoT \and Smart-* Applications \and Big Data \and Fast Data \and Benchmarks \and Performance Evaluation \and Edge Analytics \and Fog Computing}

\section{Introduction}
The proliferation of small sensors and devices that are capable of generating valuable information, has accelerated the fast development of the Internet of Things (IoT). One of the most important concepts of this paradigm is the possibility of integrating and analyzing unprecedented volumes of data, in order to make informative decisions in fields like healthcare, traffic management, water quality, air pollution and many more~\cite{iot}, by means of \emph{Smart-* Applications}. 
Smart Cities~\cite{smartcities} are the most common example of such Smart-* Applications, leveraging IoT networks to improve citizens' life. Similar to other smart environments, cities are not monolithic. They are complex integrated systems, that need to be analyzed both individually and as a whole (\textit{i.e.,} processing data across each city service to help maintain smart services). Typically, the motivation for building such smart environments is twofold: (i)
taking automated decisions in near \emph{real-time} to improve the efficiency of services and (ii) using data, computation models and machine learning for \emph{long range planning}, predictive analytics and optimization. 

Two axes are currently explored to achieve these goals: \emph{Cloud-based Big Data Analytics} and \emph{Edge Data Analytics}.

Cloud-based Big Data Analytics are used to discover new correlations and patterns, which lead to valuable insights. In this context, live data sources (e.g. camera, sensors, etc.) are increasingly playing a critical role for two reasons: firstly, they introduce an online dimension to data processing, improving the reactivity and “freshness” of the results, which can potentially lead to better insights. Secondly, processing live data sources can offer a potential solution that deals with the explosion of data sizes, as the data is filtered and aggregated, before it gets a chance to accumulate. 

Edge Data Analytics~\cite{edge} aim to deal with the cost of moving huge amounts of data across the Internet, as well as the latency overhead, which can both prevent pushing all the collected data from the sources to some centralized cloud data centers~\cite{zhang2015cloud}. The main idea is to perform an important part of the analysis at the edge of the network, located at the collection site. Such an approach allows to take local decisions and enables the real-time promise of Smart-* Applications.

Building these kinds of processing infrastructures (cloud-based, edge-based or a hybrid solution between the two) for Smart-* Applications poses new challenges at many different levels. Initially, the data has important contextual information, like spatial or temporal, that must be considered at the exact moment of storing it. There are also privacy and security concerns that can be raised, in order to protect the privacy of citizens. Also, the speed and volume of the data generated should be taken into account, as well as the heterogeneity of the different underlying sensors and systems. 

A plethora of data analytics systems have emerged to specifically deal with these aspects, especially in the Apache ecosystem (e.g. Spark~\cite{spark}, Flink~\cite{flink}, Kafka~\cite{kafka}, Storm~\cite{storm}, Samza~\cite{samza}, Pulsar~\cite{pulsar}). The wide heterogeneity in \mbox{Smart-* Applications} and data models increases even more the difficulty to compare these frameworks in terms of their features, data management functionality or performance. The "Big Data Benchmark" from Berkeley University~\cite{bbd} and the Yahoo Cloud Serving Benchmark~\cite{Cooper2010} are the most widely used benchmarks for Big Data Analytics. However, Smart-* Applications raise a set of requirements that are fundamentally different from those in traditional Big Data Applications. 

For instance, besides Big Data arriving in huge volumes (typically stored in binary large objects) and processed in batches, they also have to handle \emph{small data}, arriving at \emph{high rates}, from many \emph{geographical distributed} sources and in \emph{heterogeneous formats}, that need to be processed and acted upon with high reactivity. One could say that \mbox{Smart-* Applications} enable the possibility to detect \emph{what} is happening with a monitored object, while Big Data processing allows to understand \emph{why} this is happening.
Consequently, an open research question when choosing the right framework to process such applications is: \emph{how do existing benchmarks support and test the specific features of Smart-* Applications?}

In this article, we provide the initial answers to this question, by conducting a survey on existing benchmarks, taking into account their support for Smart-* Applications, while discussing the gap our scientific community needs to fill, in order to provide appropriate solutions. We start from two observations: (i) a key point when designing a platform for \mbox{Smart-* Applications} is the performance evaluation of the underlying systems and technologies, and (ii) appropriate benchmarks allow to effectively dimension and scale such applications. Based on our assessment, we discuss different design strategies and argue in favor of benchmark suites that leverage specific application domains.

The contributions of the article are threefold: 

\begin{itemize}
\item We characterize Smart-* Applications in terms of     \textbf{data models} and \textbf{processing              requirements};
\item We provide a survey of the current state in            \textbf{data analytics benchmarks} focusing on these     requirements;
\item We describe \textbf{a set of tools beyond             benchmarks} that can be used in the evaluation process and discuss \textbf{missing features in the context of edge analytics}, together with potential strategies towards more comprehensive benchmark suites. 
\end{itemize}

The remainder of this article is organized as follows. We first give an overview of Smart-* Applications (Section~\ref{sec:usecases}). We characterize the data models of such applications and derive a set of processing requirements, showing how these should be included in a benchmark (Section~\ref{sec:characteristics}). We present a series of benchmarks that can be used to evaluate Smart Environments, mapped to the previous requirements (Section~\ref{sec:benchmarks}). We enumerate a series of open challenges on Smart-* Applications evaluation (Section~\ref{sec:discussion}), discuss the related work (Section~\ref{sec:relatedwork}) and finally draw a conclusion on the subject (Section~\ref{sec:conclusion}). 

\section{Background: The variety of Smart-* Applications}\label{sec:usecases}

\begin{figure}[!t]
    \centering
    \includegraphics[width=12cm]{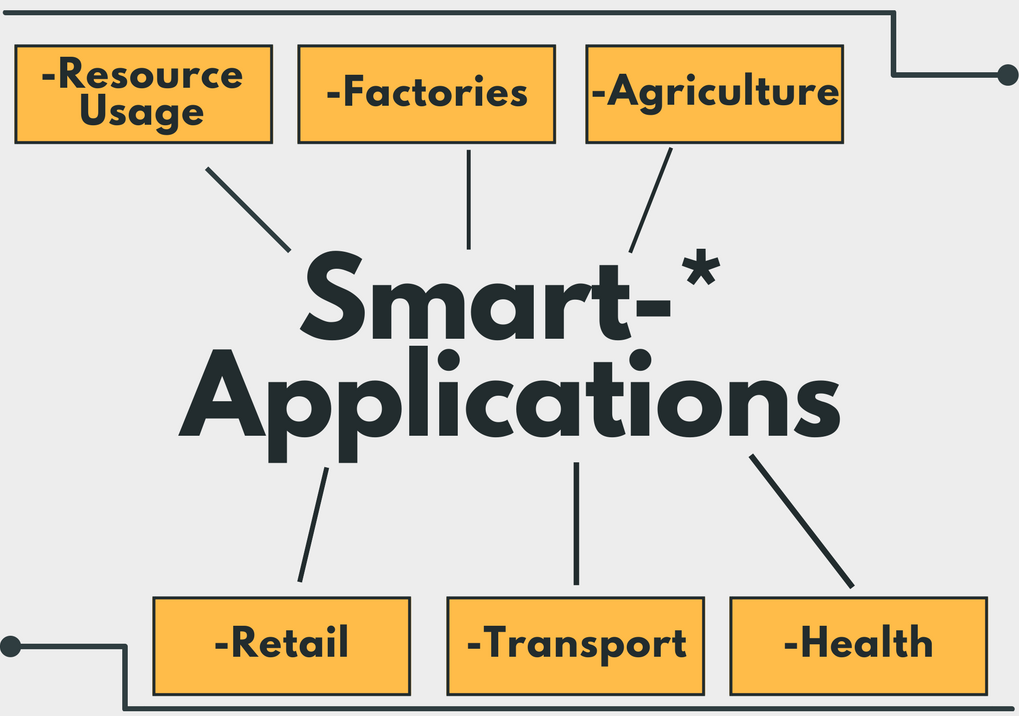}
    \caption{Smart-* Applications use cases presented in this survey. We use these domains to identify a set of characteristics inherent to data in Smart-* Applications and the challenges that arise when processing the data in this kind of environments}
    \label{fig:smart_apps}
\end{figure}

The increasing deployment of cheap IoT devices (sensors, video cameras, etc.) in all industrial or economical sectors should favor the development of Smart-* Services and Applications. This new generation of Services/Applications are expected to improve aspects of daily life and decision processes in various domains, such as resource management, factories, agriculture, transport or healthcare. 

In this section, we provide an overview of how Smart Systems and Environments could change current processes in the aforementioned domains. We show in particular that \emph{Smart Systems} and \textit{Smart Environments} are designed to work together to improve efficiency as well as reactivity. Smart systems can be considered as independent systems, that leverage IoT information to react in an autonomous manner and almost in real-time to their context. An example would be a self-driving car that processes the information coming from several sensors, in order to take decisions on the road. Smart environments bind together all these smart systems to optimize a more global goal. They can also push orders or information towards the edge. For example, they can build machine learning models, optimized with data coming from different smart systems and then push them to the edge, where critical decisions will be made. When available, we complete the description by giving information related to the amount and the frequency of the manipulated data in already implement systems and case studies. Although Smart-* Applications are still in their infancy and meant to evolve, by providing these use cases we want to draw a general picture of the particular data challenges and the diversity of these systems. 

\begin{figure}[!t]
    \centering
    \includegraphics[height=15cm]{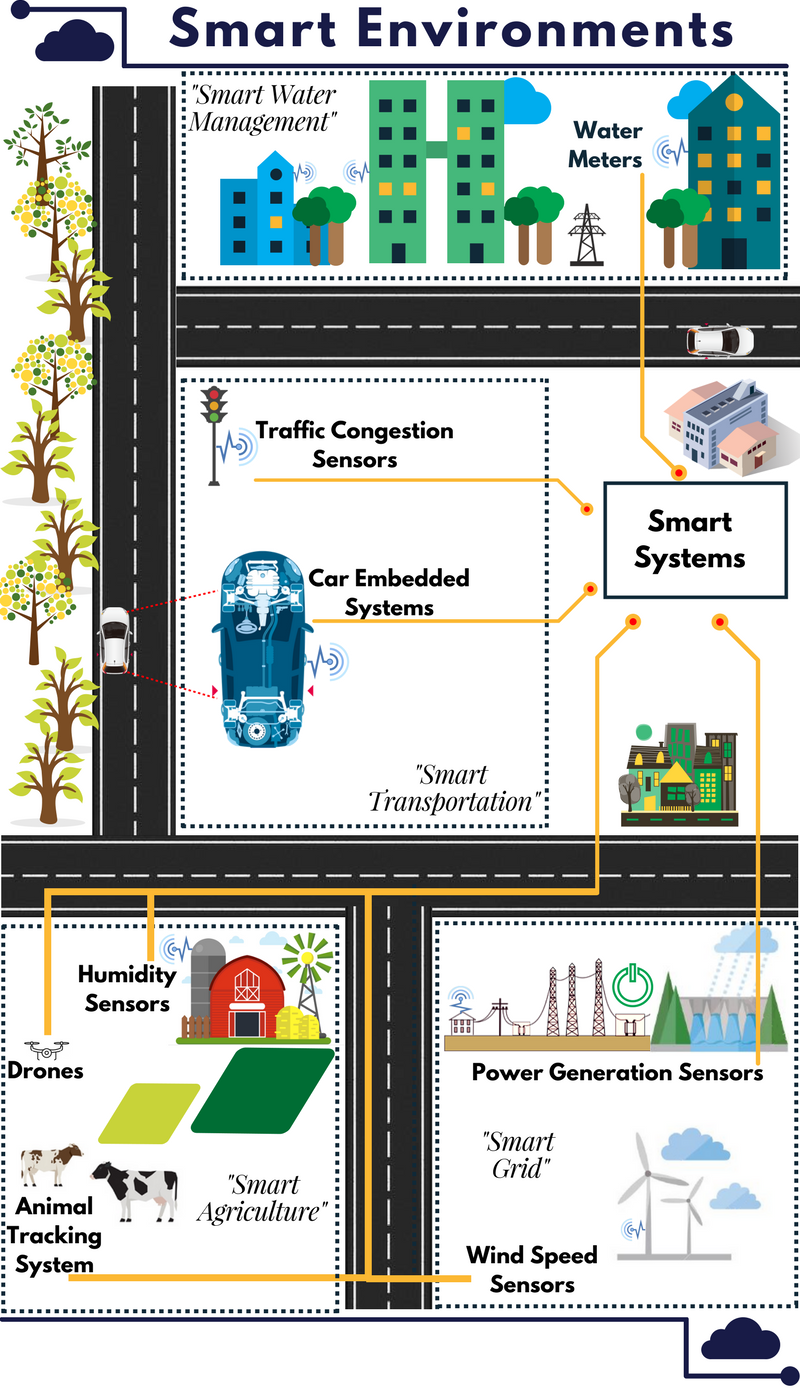}
    \caption{A graphical depiction of a set of smart environments and systems. In this case we have four different smart environments. For example, the smart agriculture environment uses an ensemble of smart systems such as drones, humidity sensors or animal tracking devices, in order to take decisions based on these measurements. Decisions can also be pushed to smart systems that behave as actuators, such as changing the traffic lights depending on the surrounding traffic or adjusting the level of water in a electricity generator dam.}
    \label{fig:smart_city}
\end{figure}

\subsection{Smart Resource Usage}\label{subsec:resource_usage}
Electricity and  water are two essential resources that keep the society running. In an ever-growing city, it is important to monitor, manage and distribute these resources efficiently, in order to maintain a sustainable growth. Smart data-gathering systems provide valuable insights for making better decisions in the management of these resources~\cite{smart-energy}. 

\smallskip

A \textit{Smart Power Grid} uses automated control converters, sensing and metering technologies, in addition to  modern energy management techniques. All these solutions enable the ability to improve the transmission efficiency of the electricity, reduce operation and management costs, and in addition, to easily integrate a smart grid with renewable energy systems, such as wind turbines and solar panels~\cite{che2017optimal}. 

The impact of equipment failures, which cause power disturbances and outages, can be largely avoided by an online power system with conditional monitoring, diagnostics and relevant protection. To this end, the intelligent monitoring and control has become essential to realize the envisioned smart grid~\cite{zheng2013smart}. The smart grid has to deal with a wide variety of data sources. One important data source is the data collected from smart meters~\cite{fang2012smart}. In a year, a smart meter which gathers data captured every 15-minutes, will eventually generate 400MB. With an expected increase of the installed smart meters, the volume of data will be in the range of petabytes or even more~\cite{lynda2014mars}.

\smallskip

\textit{Smart Water Management} is a similar smart environment, where multi-purpose meters and sensors are deployed throughout the water infrastructure and connected via wireless networks. These smart systems can send vital information back to the central system for real-time monitoring, analysis and control of water systems. For the most advanced systems, they can collect and act on information autonomously, for instance by automatically shutting off water flow and sending an alert if a leak or burst is detected~\cite{public2016managing}. Such a notification mitigates water loss and repair costs. In general, smart water meters record the water usage at 15 minute intervals~\cite{lonhouse}.

\subsection{Smart Factories}
The concept of a smart factory (a.k.a. Industry 4.0) has been proposed almost ten years ago~\cite{lucke2008smart}. 
The main idea is to move from the traditional automation process to fully connected and flexible systems, that can autonomously react in order to satisfy the customer demands, according to the states and the events that can occur on the different production systems. While planing and scheduling techniques have been used for a while now, in order to optimize the production yields, the use of IoT sensors/cameras will allow manufacturers to go one step ahead, by controlling the whole production chain in a more efficient way.

For instance, dedicated smart systems will be in charge of monitoring the wear of mechanical engines, taking into account environmental conditions (temperature, humidity, contaminants, etc.). Leveraging this data, they will be able to autonomously program some maintenance operations, before any failures occur. These events will be integrated in the global environment, being in charge of orchestrating all systems, including also operators and engineering planning. Furthermore, the operation management system will integrate additional external system events, coming from the customers who made the command orders to the suppliers that will deliver the raw materials. All these exchanges will lead to a constant stream of data, that will provide valuable information for manufacturers to optimize their process and thus reduce the costs and the waste.

The form of the required data and how it will be manipulated has not been well identified yet. Envisioned scenarios include massive data exchanges that should be uploaded to and processed by cloud computing systems to ensure inter-coordination~\cite{wang2016implementing}. However, to be able to absorb such a deluge of data, it will be mandatory to perform as much data analysis as possible close to the smart systems (i.e. at the edge of the global system).

\subsection{Smart Agriculture}\label{subsec:agriculture}

Monitoring and gathering data from a farm benefits both the farmer and its consumers~\cite{Libelliumag}. The data comes from an IoT infrastructure, where sensors collect soil parameters, ambient and weather conditions, chemical components and even fruit size. These sensors send data with different levels of granularity, depending on the required speed. For example, ambient temperature does not change as often as the position of one animal that needs to be tracked~\cite{linklabs}. Another factor is the level of precision needed, as the number of sensors per acre of land depends on the level of detail required.

Because crop growth is a slow process and environmental conditions do not change as fast as in previously mentioned scenarios, data velocity will not be the greatest challenge. In most envisioned use cases, the sampling rate is more than 5 minutes~\cite{Libelliumag,jayaraman2016internet}. Some other scenarios involve mobile sensors, installed in tractors or air balloons, that upload their data in a batch manner, as soon as they gain connectivity through a gateway~\cite{perera2014sensing}. 
But even if there is not a high speed streaming requirement, smart agriculture is a complex process in terms of data management, where several data insights coming from different sources have to be combined, normally through semantic web and linked data techniques~\cite{jayaraman2016internet, Sonka2014}. For instance, local weather should be predicted in time, in order to act in advance against frost. In addition, data from different farms needs to be combined to build accurate models, since a single data source might not be enough. Consequently, the complexity of Smart Agriculture mostly lies in linking all of these heterogeneous data and the analytics required.

\subsection{Smart Retail Store}\label{subsec:retail}

The use of Smart Systems/Environments has been also recognized to be important for the future of retail~\cite{smart_retail_2014}. Retailers create and strengthen relationships with their customers through individual pricing, subscription plans and also evolving the salespeople job profiles. From the customer side, the factors changing the game are service access, shopping experience and finally personalization. 

Data sources are not only limited to transaction systems, which provide past user purchases, returns or changes. Most recent data sources are augmented by real-time data, representing customer retail interaction. For instance, the new generation of retail shops shows that an optimal product placement can be arranged through real-time customer tracking and analysis of their behavior~\cite{intel_levis}. In this task, devices such as proximity sensors, radio-frequency identification systems and also retail cameras are considered critical sources of data. Moreover, customer activities such as gazing, focusing and choosing products inside a store are captured as data traces~\cite{retail_research_glasses}. Also, enhancement of customer experience can be achieved using mobile applications with augmented reality. In this case, customer satisfaction grows in pair with more efficient shopping~\cite{retail_augmented_reality}.

Already implemented Smart Retail Stores, such as the Amazon Go~\cite{amgo}, demonstrate the data processing challenges associated with these systems. Through the use of deep learning techniques, computer vision and sensor fusion technology, customers can enter the shop, pick up the products they want and walk out without paying at the cash register in the end. A receipt arrives in the customer's e-wallet and everything is paid electronically. This is a clear example of a Smart Environment, in which data from different sources has to be combined to provide fast decisions and a seamless service to customers.

\subsection{Smart Transportation} \label{subsec:transportation}

Smart Transportation gives users smarter choices when deciding on the route guidance in the cities. Also, it helps reducing traffic congestion and fighting pollution. The navigation system in each vehicle, combined with live traffic data and parking information, can calculate the "best route" in terms of fuel efficiency and time-saving. The live traffic data may come from the network of connected smart traffic lights, security cameras as well as other vehicle GPS devices. The data can be used to plan and improve public transportation systems and reduce traffic congestion.

More specifically, we distinguish Smart Transportation Systems and Smart Transportation Environments. Intuitively, Smart Systems are autonomous cars, short-term rent electric vehicles or an autonomous subway. The environments are smart traffic management, smart fleets and smart city infrastructure management. One of the most critical environments for citizens is an intelligent public transport system. 

The challenge of Smart Transportation systems is the amount of data generated by different sensors and equipment in the cities. Even simple solutions generate enormous amounts of data for analysis, e.g. 2GB generated for each of the 26,000 loop detectors under the highways in the State of California~\cite{varaiya2005reducing}. Not only the data comes from a broad spectrum of devices, but also at high velocity to capture live events on the roads. These challenges put the analytic engines, communication and storage systems on a great deal of stress.

\subsection{Smart Health}\label{subsec:health}

Smart Health (sHealth) promises to revolutionize the healthcare sector by using the context-aware network and sensing infrastructure~\cite{solanas2014sHealth}. For instance, a representative example of an sHealth environment is the continuous monitoring of vital signs~\cite{Konstantas2004}, which consists of patients with heart disease wearing bracelets that monitor their vital constants. In case of a heart attack, the bracelet detects the fluctuations in the blood pressure and heart rate and sends an alert to the city health services. After receiving the alert, the traffic conditions are analyzed, as described in Section~\ref{subsec:transportation}, and an ambulance is dispatched through the best possible route, demonstrating how the whole context around the data source is also included in the decision-making process.

sHealth environments face all three of the traditional Big Data challenges: variety, velocity and volume~\cite{solanas2014sHealth}. In the example above, monitoring vital signs with an interval of 1 minute can generate up to 2GB of data per month per patient~\cite{Albright2011}, which easily translates to terabytes of data in a Smart City (volume). The collected data is also very diverse and needs to be integrated with the city health system (variety), and finally, it needs to be analyzed in near-real-time to provide a useful service to citizens (velocity). Moreover, given the highly sensitive nature of health data, novel security and privacy measures are needed to guarantee the authenticity of the data and protect the citizens privacy~\cite{chung2015PGHD}.

On the other hand, an example of a Smart Health system is the haptic controller increasingly used to perform surgeries and treatments from a distance (Remote Surgery or "telesurgery") by applying forces, vibrations and motions to recreate the movements and touches of the surgeon~\cite{Marescaux2002}. In this case, the volume of generated data does not pose a challenge but the velocity requirement is extremely important as the surgeon's motions must be analyzed and replicated by the controller in real-time to guarantee the success of the whole procedure~\cite{Butner2001,Smithwick1996}.

\section{Key requirements of Smart-* Applications}\label{sec:characteristics}

From the previously presented Smart Environments, systems and their related case studies, we can identify a set of common data characteristics. These characteristics are inherently related to the different ways users can process and store this data. For example, to store measures that are periodically sent by sensors, it would make sense to use a time series database~\cite{xu2014tsaaas}, whereas for video images, object storage is widely used \cite{rodriguez2012video}.

When evaluating and defining benchmarks for any of these Smart-* Applications, we need to take into account both the characteristics of the data and its process/storage requirements, since they will drive the design choices of the systems supporting them. In this section, we provide a list of both the aforementioned aspects. The objective is to have a clear definition of features that a benchmark should fulfill to evaluate a data analytics system in this context. The relation between the data characteristics and the processing requirements does not have to follow an one-to-one relation. In some cases, some particular types of data might need one or more types of processing, depending on the use case. For example, small and fast data coming from sensors do not always have to be processed in real-time, but they could also be processed in a batch manner at a later stage.

\subsection{Data characteristics}\label{subsec:datacharacteristics}

\begin{figure}[!t]
    \centering
    \includegraphics[width=12cm]{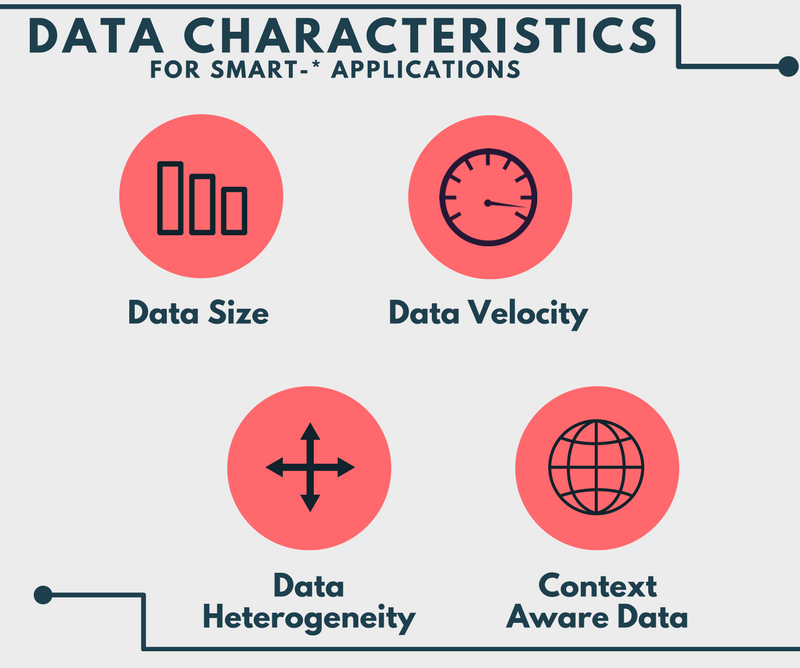}
    \caption{Data characteristics of Smart-* Applications}
    \label{fig:data-characteristics}
\end{figure}

Smart-* Applications are fueled by data that provides a fine-grained, wide and real-time understanding of the environments they manage. The nature of this data changes depending on the environment. Leveraging the use cases described in Section~\ref{sec:usecases}, we have identified a series of data characteristics (see Figure~\ref{fig:data-characteristics}) that any benchmark designed to evaluate data analytics for Smart-* Applications should have. In this section, we provide a comprehensible list of these characteristics. 

\subsubsection{\textbf{Data Size}} Primarily, the size of the data generated by Smart-* Applications depends on the system or  environment: the amount of considered elements (number of sensors, users etc.), data type (video, images, temperature measures etc.) and acquisition frequency (every second, minute, hour etc.). Furthermore, we can distinguish between the size of the historical data that is stored in repositories and the size of the incoming streamed data, which is linked to the data arrival rate aspect.

The size of archived data depends on the lifetime of the system and the importance of data. During the lifetime of the system, data is archived and can be down-sampled. For instance, some per-minute sensor measurements can be aggregated into per-hour measurements in order to reduce storage space. In contrast, data sets with valuable data, such as images and videos, are stored in petascale storage systems. These data is usually used not only for analysis, but also for training and improving machine learning models, e.g. Tesla acquiring data from thousands of vehicles supplied with specific autonomous car equipment~\cite{TESLA_STORAGE}. 

Regarding stream processing, data volume can vary depending on the use case. We mentioned before how current deployments of Smart Agriculture sensors do not generate that much data, compared to magnitudes of terabytes per second for a million people population utilizing smart grid~\cite{smart_grid_data_sizes_2014}.

It is essential for a benchmark to provide different settings for the volume of the data they generate. This enables the valuation of the scalability of the system as the data size grows.

\subsubsection{\textbf{Data Velocity}} In the same vein of Big Data characteristics, velocity will also introduce a set of requirements for data processing in Smart-* Applications. This can be measured in terms of data size per second (e.g. GB per second). If data consists of records of a fixed size, it can also be measured in terms of records per second. Data can arrive at different rates and systems should be stressed under these conditions for a proper evaluation. This rate can vary depending on the use case. For example, in a Smart Health application we can have rates up to 25,000 records per second~\cite{cortes2015stream}, while the Smart Agriculture use case could contain much slower data. Another distinctive feature is that the data arrival rate can be tied to certain events, like commuters going through the turnstile of a metro station, and so it can be variable. Benchmark users should be able to change the data generation rate and its patterns to create different scenarios and have guarantees that the systems they design will work in the real world.

\subsubsection{\textbf{Data Heterogeneity}} One of the key aspects of Smart-* Applications is that they aim to model the real world, in order to take better decisions. This involves a sensing layer~\cite{sun2016internet} with data coming out of different domains and representing diverse information. We can have geolocation data coming from mobile vehicles, time series data coming from sensors or video images from cameras~\cite{li2015big,hossain2017cloud}. The implications of this in a benchmarking process for data analytics are twofold. Firstly, it changes the way of storing and processing the data. In the time series example, the data access will have to consider different dimensions, like timestamp or sensor, while the geolocation data of a GPS system might be used by a graph processing engine to find the shortest path between two points. Secondly, this data might not have a common format, since it comes from different sources. Sometimes it does not even have a format and it comes in an unstructured way, such as text or images. This is a common feature for Smart-* Applications where data needs to be integrated and preprocessed in order to extract meaningful insights.
A benchmarking tool should consider generating or including datasets that capture this heterogeneous nature.

\subsubsection{\textbf{Context-aware Data}}

Smart environments are based on their ability to sense the world that surrounds them and reason on these observations in order to act accordingly \cite{cook2009ambient}. Context-aware data is of great importance when trying to reason about the current situation of an environment and an area of research on its own \cite{perera2014context}, with the definition for "context" widely accepted as "any information that can be used to characterize the situation of an entity". Some examples of context include the location (longitude, latitude, etc.), the identity (sensor Id, user Id, etc.) or the time or physical events which are normally recorded by the sensor itself (movement, temperature, humidity, etc.). For instance, in the aforementioned smart transportation use case, a unit of context for a car could be its location, its speed and the current time. This information has to be included in the data used, as it will be necessary to evaluate any data access patterns or algorithms that use such contextual knowledge. 

\subsection{Processing and storage requirements}\label{sec:processingrequirements}

\begin{figure}[!t]
    \centering
    \includegraphics[width=12cm]{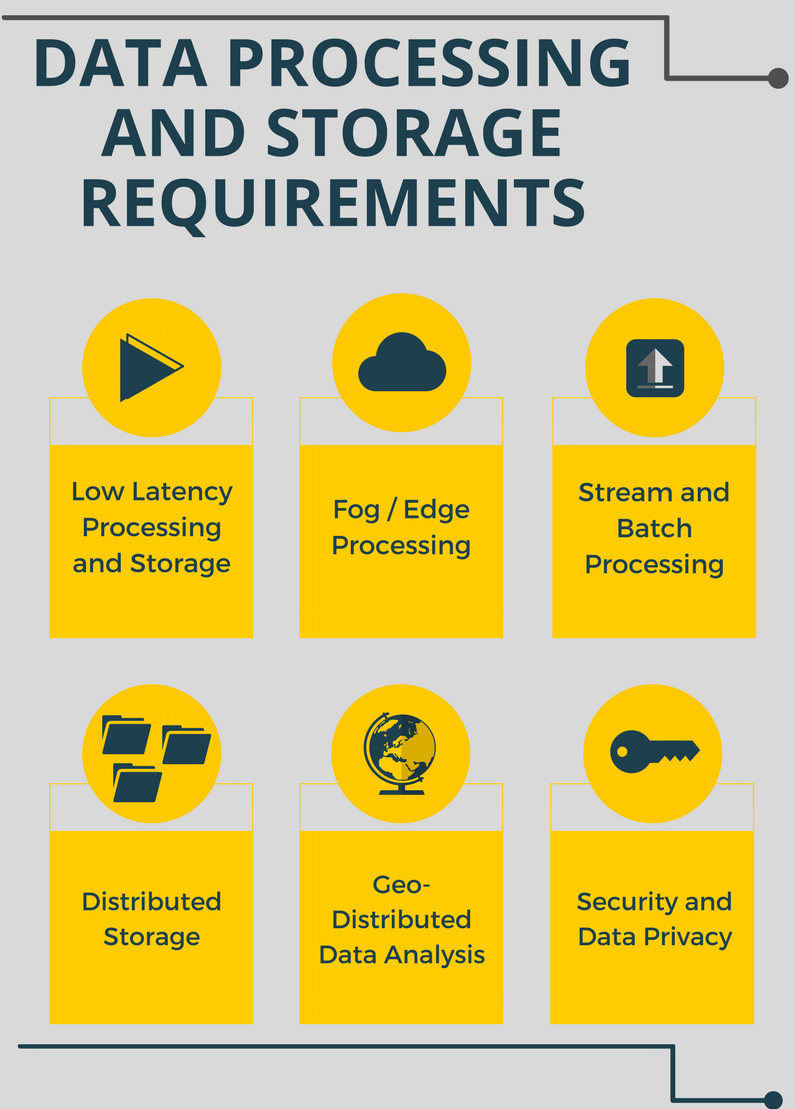}
    \caption{Data processing and storage requirements of Smart-* Applications}
    \label{fig:process-requirements}
\end{figure}

In the previous subsection, we have focused on understanding the features of the data handled by Smart-* Applications. Below, we list a set of requirements needed to process and store all of this data (see Figure~\ref{fig:process-requirements}). 

\subsubsection{\textbf{Low-latency processing and storage}}

Smart-* Applications and low latency are heavily linked. The potential value that can be extracted from raw data is immense, but for it to express its full potential, one must be able to analyze and derive insight from it in almost real-time. The network, processing and storage latency impact each level of a stack where data is generated from physical events, processed and stored and finally used to extract knowledge from it. Each of these levels has a significant impact on time-to-decision: the time it takes to trigger alerts, deliver them to a relevant end point and then to act accordingly. The longer the time-to-decision, the more the value of the information is diminished.

In this respect, a benchmarking tool should be able to measure the time it takes to perform any of the aforementioned processing, ensuring that the critical latency requirements of Smart-* Applications can be met.

\subsubsection{\textbf{Fog/Edge processing}}

\begin{figure}[!t]
    \vspace{-1mm}
    \includegraphics[width=12cm]{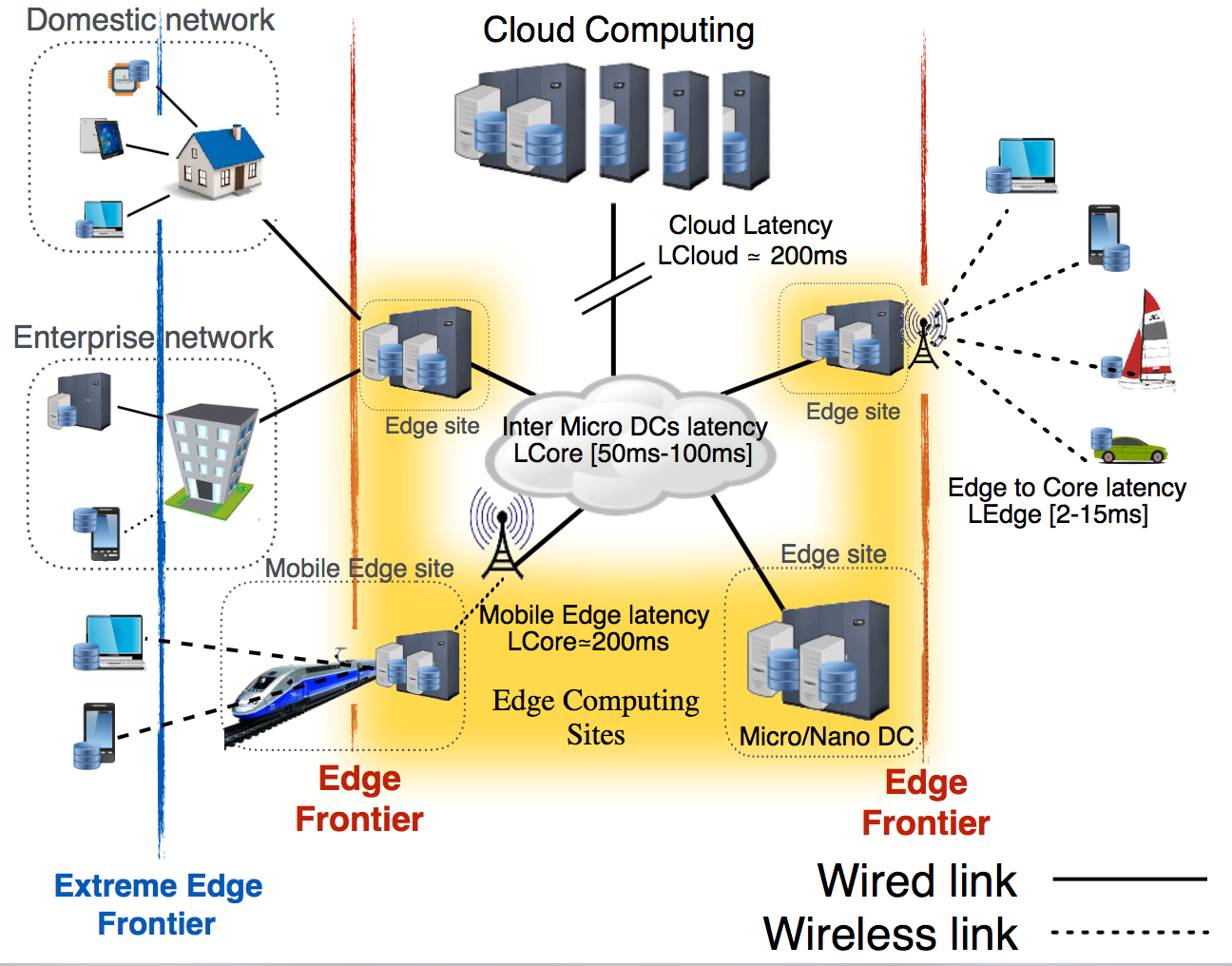}
    \centering
    \caption{Fog/Edge Computing Infrastructure Overview\cite{edgecomputing}}
    \label{fig:fogcomputing}
    \vspace{-1mm}
\end{figure}

Edge and Fog processing are born from the necessity of satisfying the latency requirement explained in the previous paragraph. In this computing paradigm, processing facilities are moved near the data sources, significantly reducing time-to-reaction and transfer latency, triggering events as soon as possible as well as pre-aggregating data for further processing. A general overview of the infrastructure involved in edge computing can be seen in Figure~\ref{fig:fogcomputing}. 

Edge has several unique characteristics in contrast to the already well-known cloud, particularly those related with infrastructure. To start off, devices that support this type of processing can be mobile, since they can reside in vehicles such as trains or cars, and consequently they can lose connectivity if they reach poor signal areas. Moreover, they can consist of low power devices that can be either switched off or run out of battery. All of these create an unstable infrastructure, where fault tolerance must be ensured for any application or system running on top. Another important feature is that the devices at the Edge can be resource-constrained in terms of memory, storage, energy or computational power. Smart-* Applications and the systems supporting them must take this into consideration, by providing lightweight computation and low system requirements.

These distinctive features of edge computing have to be included in our evaluation process. By using key indicators, such as fault tolerance or resource usage, we can ensure that Smart-* Applications can be deployed in a real Fog/Edge environment with guarantees.

\subsubsection{\textbf{Stream and Batch Processing}}
With the advent of the IoT sensors and devices, the development of Smart-* Applications is expected to exacerbate the amount of data flowing from all connected objects to private and public cloud infrastructures. 
Batch processing is still used to complement this online dimension with a machine / deep learning dimension and 
gain more insights based on historical data. The ultimate goal is to have an online/real-time front-end for processing in the edge, close to the location where data streams are generated, while the cloud will only be used for off-line back-end processing, mainly dealing with archival, fault tolerance and also further processing that is not time-critical.

\subsubsection{\textbf{Distributed Storage}}


The amount of data generated by Smart-* Applications far exceeds the storage capacities of a single machine. Storing such vast amounts of data is challenging. At the scale of a Smart City potentially leveraging millions of sensors, the associated challenges are even higher. Yet, storing the data is a requirement for many applications. When the data is processed as streams, storage can serve as a buffer for coping with data bursts, when the data generation rate exceeds the analytics resources. When the data is processed as batches, it serves as a data lake, storing the raw data awaiting to be processed. Finally, on the edge of a sensor network with intermittent connectivity, it enables data to be buffered, while awaiting to be transferred to a remote data repository or analytics facility.

Being able to benchmark storage systems handling Smart-* Applications data is important. Indeed, for all the aforementioned use cases, ensuring that the storage system is scaled appropriately is critical. It should be able to handle writes at the peak data generation velocity, in order not to lose precious incoming data. Also, it should provide the required read performance for the data analytics systems in order not to cause data starvation and waste valuable computing resources.

Depending on the precise requirements of the application, multiple storage paradigms can be considered. For instance, object stores are ideal for storing the raw data originating from the sensors (Ceph~\cite{ceph}, OpenStack Swift~\cite{swift}, Týr~\cite{tyr}), while columnar stores are best suited for structured data (BigTable~\cite{bigtable}, Cassandra~\cite{cassandra}). A large, complex application may even rely on different storage paradigms for distinct data processing stages. Multiple storage paradigms imply that different benchmarking tools must be leveraged depending on the precise system used and its associated performance requirements. For complex use-cases relying on multiple storage systems, a combination of benchmarks may be necessary. As such, choosing the relevant benchmarks is essential for adequately scaling and validating the storage requirements for any large-scale Smart-* Application. 

\subsubsection{\textbf{Geo-distributed data analysis}} 
As the data gathered from many sensors could possibly be stored in datacenters around the globe, some limitations could arise on the communication between these datacenters. Data analytics that exploit all the data at the same time, in order to derive meaningful results, could hinder on bandwidth limits, regulatory constraints, problematic runtime environments or monetary costs~\cite{zhang2018towards}. Computation procedures could be distributed, in order to take advantage of more than one datacenter~\cite{Hung2015} and improve the job completion time. 

The latency and bandwidth limitations through the weak links that connect the different datacenters, has to be taken into account in the evaluation process, either by using physically distributed computing resources, such as the different Amazon EC2 sites, or by emulating these constraints. Moreover, the data generation part of the benchmark should consider this property and perform the data placement based on some user-selected geo-distributed criteria for this scenario.

\subsubsection{\textbf{Security and Data Privacy}}
Aside from the inherent security challenges that arise when using IoT devices~\cite{xu2014security}, it is critical to ensure the privacy of the data in Smart-* Applications. Much of it comes from user devices or sensors with confidential information, such as geographical locations, energy/water consumption or health conditions. This is especially necessary when the computation assets, which store and process the data, are hosted in the cloud and can be easily compromised. For this reason, several design solutions for privacy-aware smart cities have been proposed~\cite{Li2016,Martinez2013}. Other issues, that are not exclusive to Smart-* Applications but still need to be taken into consideration, include identification and authentication~\cite{Riahi2013}, guaranteeing the integrity of data~\cite{Mahmoud2015} and minimizing software vulnerability and malware in connected IoT devices~\cite{Zhang2014}.

\section{Benchmarking Smart-* Applications}\label{sec:benchmarks}

\begin{table*}
\scriptsize
\centering
\caption{Benchmarks and the different workloads they provide, software stacks they support, metrics they offer as well as the datasets used by each workload}
\label{table_label_benchmarks_desc}


\makebox[\textwidth][c]{\begin{tabular}{|p{2.1cm}|p{2.2cm}|p{2.8cm}|p{3.6cm}|p{4.8cm}|}
\hline
\rowcolor[HTML]{EFEFEF} 
\textbf{Benchmark}                      & \textbf{Workloads}                      & \textbf{Software stacks}                 & \textbf{Metrics}                                               & \textbf{Datasets}                               \\ \hline
                                        & Machine learning                        & Mahout                                   & Latency                                                        & Wikimedia dataset                               \\ \cline{2-5} 
                                        & Data caching                            & Memcached                                & Throughput, Latency                                            & Twitter dataset                                 \\ \cline{2-5} 
                                        & Data serving                            & YCSB with Cassandra                      & Throughput, Latency                                            & Synthetic dataset                               \\ \cline{2-5} 
                                        & Graph Analytics                         & Apache Spark - Page Rank                 & Latency                                                        & Twitter dataset                                 \\ \cline{2-5} 
                                        & In-memory Analytics                     & Apache Spark - ALS                       & Latency                                                        & Movie reviews dataset                           \\ \cline{2-5} 
                                        & Video streaming                         & Nginx                                    & Throughput                                                     & Synthetic dataset                               \\ \cline{2-5} 
                                        & Web search                              & Apache Solr                              & Throughput                                                     & Indexed crawled websites                        \\ \cline{2-5} 
\multirow{-8}{*}{\textbf{CloudSuite}}   & Web serving                             & Elgg, MySQL                              & Throughput                                                     & N/A                                             \\ \hline
\textbf{LinearRoad}                     & Stream Processing                       & Technology agnostic                            & Response time, query accuracy                                  & Traffic dataset                                 \\ \hline
                                        &                                         &                                          &                                                                & CITY (smart city sensors)                       \\ \cline{5-5} 
                                        &                                         &                                          &                                                                & NYC Taxi Cab (taxi trips)                       \\ \cline{5-5} 
                                        &                                         &                                          &                                                                & GRID (power meters)                             \\ \cline{5-5} 
\multirow{-4}{*}{\textbf{RioTBench}}    & \multirow{-4}{*}{Stream Processing}     & \multirow{-4}{*}{Apache Storm}           & \multirow{-4}{*}{\shortstack[l]{Latency, throughput, jitter,\\resource usage}}  & MHealth (sensors from patients)                 \\ \hline
                                        & Machine Learning                        &                                          &                                                                &                                                 \\ \cline{2-2}
                                        & Graph Analytics                         &                                          &                                                                &                                                 \\ \cline{2-2}
\multirow{-3}{*}{\textbf{SparkBench}}   & OLTP                                    & \multirow{-3}{*}{Apache Spark}           & \multirow{-3}{3.6cm}{Latency, data process rate (Mb/sec)}          & \multirow{-3}{*}{Synthetic}                     \\ \hline
                                        &                                         &                                          &                                                                & Vehicle traffic (traffic sensors)               \\ \cline{5-5} 
                                        &                                         &                                          &                                                                & Parking dataset (parking sensors)               \\ \cline{5-5} 
                                        &                                         &                                          &                                                                & Weather dataset (weather sensors)               \\ \cline{5-5} 
                                        &                                         &                                          &                                                                & Pollution dataset (pollution sensors)           \\ \cline{5-5} 
                                        &                                         &                                          &                                                                & User location (geolocation of mobile users)     \\ \cline{5-5} 
\multirow{-6}{*}{\textbf{CityBench}}    & \multirow{-6}{2.2cm}[4pt]{RDF Stream Processing} & \multirow{-6}{2.8cm}{RSP Engines (C-SPARQL)} & \multirow{-6}{3.6cm}{Latency, memory consumption}                  & Cultural dataset (planned cultural city events) \\ \hline
\textbf{BigBench}                       & OLTP                                    & Technology agnostic                            & Latency                                                        & E-commerce: web logs, item reviews and sales    \\ \hline
\textbf{IoTABench}                       & OLTP                                    & Technology agnostic                            & Latency                                                        & Synthetic data (smart grid)                     \\ \hline
\textbf{YCSB}                           & Data Serving                            & Technology agnostic                            & Throughput, Latency                                            & Synthetic data                                  \\ \hline
                                        & Graph Analytics                         & Hadoop, Spark, Flink, MPI                &                                                                & Google web graph, Facebook network              \\ \cline{2-3} \cline{5-5} 
                                        & Batch processing                        & Hadoop, Spark, Flink, MPI                &                                                                & Wikipedia dataset                               \\ \cline{2-3} \cline{5-5} 
                                        & Stream processing                       & Spark, JStorm                            &                                                                & Synthetic                                       \\ \cline{2-3} \cline{5-5} 
                                        & OLTP                                    & HBase                                    &                                                                & ProfSearch Resumes                              \\ \cline{2-3} \cline{5-5} 
                                        & Machine Learning                        & Hadoop, Spark, Flink, MPI                &                                                                & Movie Reviews                                   \\ \cline{2-3} \cline{5-5} 
                                        & Data Warehouse                          & Shark, Hive, Impala                      &                                                                & E-commerce Transaction Data                     \\ \cline{2-3} \cline{5-5} 
\multirow{-7}{*}{\textbf{BigDataBench}} & Multimedia Analytics                    & MPI                                      & \multirow{-7}{*}{\shortstack[l]{Latency, throughput,\\data processed (Mb/sec)}} & ImageNet, MNIST, Audio files,                   \\ \hline
                                        & Batch processing                        & Hadoop, Spark                     &                                                                &                                                 \\ \cline{2-3}
                                        & Stream processing                       & Storm, Spark, Flink, Gearpump                                   &                                                                &                                                 \\ \cline{2-3}
                                        & Machine learning                        & Hadoop, Spark                     &                                                                &                                                 \\ \cline{2-3}
\multirow{-4}{*}{\textbf{HiBench}}      & Graph analytics                         & Hadoop, Spark                     & \multirow{-4}{3.6cm}[5pt]{Latency, throughput, resource usage}          & \multirow{-4}{*}{Synthetic}                     \\ \hline
\end{tabular}}
\end{table*}

In this section, we describe a set of benchmarks that satisfy some of the requirements explained in Section~\ref{sec:characteristics}. Each subsection contains a brief explanation of the benchmark and the use case(s) for which it has been designed. A comprehensible list of them is provided in Table \ref{table_label_benchmarks_desc}, where we detail the name of the benchmark, the different workloads they provide, the software stacks they currently support, the metrics that form the output of the benchmark and the type of datasets they include. Note that, if the benchmark does not have an associated software stack, it is indicated as \textit{"technology agnostic"}.

\subsection{CloudSuite~\cite{ferdman2012clearing}}


CloudSuite is a benchmark suite for performance evaluation of cloud services. It includes eight benchmarks that represent massive data manipulation with tight latency constraints. These workloads have been selected based on their popularity in modern datacenters.  They are divided in two groups: \textit{(i) offline benchmarks} and \textit{(ii) online benchmarks}. The former is related to data analytics, recommendation systems and graph analytics. It operates on large datasets, while the completion time is the major metric of this group. The latter is  related to data serving, data caching, media streaming, web serving and web search. It also processes large datasets, but the major metric is the latency of requests under QoS objectives.

\subsection{Linear Road~\cite{Arasu2004}}
Linear Road simulates a toll system for the motor vehicle expressways of a large metropolitan area. The benchmark specifies a variable tolling system for a fictional urban area, including accident detection and alerts, traffic congestion measurements, toll calculations and historical queries. Linear Road is designed to measure how well a system can meet real-time query response requirements in processing high volume streaming and historical data.  This benchmark is suited to emulate smart transport use-cases.
%
%

\subsection{RIoTBench~\cite{ShuklaCS17}}
The \textbf{R}eal-time \textbf{IoT} \textbf{Bench}mark suite has been developed to evaluate Data Streaming Processing Systems (DSPS) for streaming IoT applications. The benchmark includes 27 common IoT tasks, classified across various functional categories and implemented as reusable micro-benchmarks. It further proposes four IoT application benchmarks composed from these tasks, that leverage various dataflow semantics of DSPS. The applications are based on common IoT patterns for data pre-processing, statistical summarization and predictive analytics. These are coupled with four stream workloads sourced from real IoT observations on smart cities and fitness, with peak stream rates ranging from 500 to 10,000 messages/second and diverse frequency distributions. DSPSs are key elements in the IoT context. RIoTBench can be used to evaluate them, especially in analyzing and making real-time decisions based on data coming from multiple sources.
\subsection{SparkBench~\cite{Li2015}}
SparkBench is a benchmarking suite specifically for Apache Spark. It comprises of a representative and comprehensive set of workloads, belonging to four different application types that are currently supported by Apache Spark, including machine learning, graph processing and SQL queries. The chosen workloads exhibit different characteristics that exhibit several system bottlenecks: CPU, memory, shuffle and IO. Hence, SparkBench covers three categories of algorithms widely used in Smart-* Applications and can be used to evaluate these algorithms.

\subsection{CityBench~\cite{ali2015citybench}}
This benchmark addresses a special type of streaming engines, called RDF Stream Processing (RSP)~\cite{le2011native}. The importance of these kind of engines is obvious, especially if we consider the data heterogeneity characteristics of Smart-* Applications, where the Semantic Web and Linked Data can play a relevant role. The benchmark uses real-time datasets, gathered from sensors installed for Smart Cities applications, including weather, pollution or parking datasets, to name a few. CityBench allows changing the input streaming rate, it provides different dataset sizes and supports different data access patterns, which align with the requirements we enumerated in Section~\ref{sec:characteristics}. This benchmark aims to fill the existing gap on querying linked datasets. The main metrics included are the latency and the memory consumption, for the several queries over the different state-of-the-art RSP engines.

\subsection{BigBench~\cite{Ghazal2013}}
BigBench is an end-to-end Big Data benchmark proposal based on a fictitious retailer who sells products to customers via physical and online stores. The proposal covers a data model, a data generator and workload descriptions. It addresses the variety, velocity and volume aspects of Big Data systems, containing structured, semi-structured and unstructured data. Additionally, the data generator provides scalable volumes of raw data based on a scale factor. This allows BigBench to be used to test the batch processing aspects of Smart-* Applications.

\subsection{IoTABench~\cite{Arlitt2015}}
The \textbf{IoT} \textbf{A}nalytics \textbf{Bench}mark (IoTABench) aims to bridge the divide between Big Data research and practice, by providing practitioners with sensor data at production scale. The initial implementation of IoTAbench focuses on a single use case, specifically the smart metering of electrical meters, but the benchmark can be easily extended to multiple IoT use cases. The benchmark can generate a large dataset (22.8 trillion distinct readings, or 727 TB of raw data) with realistic properties using a Markov chain-based synthetic data generator. Hence, this benchmark is ideal for evaluating real-life Smart-* Applications, that are based on sensor data.

\subsection{YCSB~\cite{Cooper2010}}
The Yahoo! Cloud Serving Benchmark (YCSB) is an open-source and extensible benchmark, that facilitates performance comparisons of cloud data serving systems. It models systems, based on online read and write access to data, while it provides a set of workloads that generate client requests to different back-ends like Cassandra, RAMCloud, HBase or MongoDB. YCSB is easily customizable and allows end-users to define new types of workloads, to configure the clients with different parameters as well as to add new back-ends. YCSB can be used to evaluate cloud services, which are an essential part of Smart-* Applications, ranging from open data access APIs to use cases related to health, social networking and public services.
\subsection{BigDataBench~\cite{Wang2014}}
BigDataBench is an open-source, multi-discipline benchmark, coming from common research and engineering efforts from both industry and academia. The current version of the benchmark models five typical and important Big Data application domains: search engines, social networks, e-commerce, multimedia analytics and bioinformatics. In total, it includes 14 real-world data sets and 34 Big Data workloads. Moreover, it provides an interesting set of workloads in both stream and batch implementations.
\subsection{Intel HiBench~\cite{Huang2011}}
HiBench is a realistic and comprehensive benchmark suite, which consists of a set of programs including both synthetic micro-benchmarks and real-world applications. The benchmark contains a set of workloads for Hadoop and Spark, as well as streaming workloads for Spark Streaming \cite{spark}, Flink \cite{flink}, Storm and Gearpump. It helps evaluate different Big Data frameworks in terms of speed, throughput and system resource utilization on a set of workloads relevant to several Smart-* Applications.

\subsection{Summary}

\begin{table}[!t]
\centering
\scriptsize
\newcommand{\STAB}[1]{\begin{tabular}{@{}c@{}}#1\end{tabular}}
\begin{tabular}{|c|c|l|l|l|l|l|l|l|l|l|l|}
\hline
\rowcolor[HTML]{EFEFEF} 
             & \multicolumn{5}{c|}{\cellcolor[HTML]{EFEFEF}Data Characteristics}                 & \multicolumn{6}{l|}{\cellcolor[HTML]{EFEFEF}Data Processing and Storage Requirements}                                                   \\ \hline
\rowcolor[HTML]{EFEFEF} 
             & \rotatebox[origin=c]{90}{Data Size} & \rotatebox[origin=c]{90}{Data velocity} & \rotatebox[origin=c]{90}{Context-aware data} & \rotatebox[origin=c]{90}{Data heterogeneity} & \rotatebox[origin=c]{90}{Low latency} & \rotatebox[origin=c]{90}{Fog/edge processing} & \rotatebox[origin=c]{90}{Stream processing} & \rotatebox[origin=c]{90}{Batch processing} & \rotatebox[origin=c]{90}{Distributed storage} & \rotatebox[origin=c]{90}{ Geo-distributed data analysis  } & \rotatebox[origin=c]{90}{Security and privacy} \\ \hline
\cellcolor[HTML]{EFEFEF}CloudSuite   & \textbf{} & \checkmark             &                    &                    & \checkmark           &                     & \checkmark                 & \checkmark                & \checkmark                   &                               &                      \\ \hline
\cellcolor[HTML]{EFEFEF}LinearRoad   & \checkmark         &               & \checkmark                  &                    & \checkmark           &                     & \checkmark                 &                  & \checkmark                   &                               &                      \\ \hline
\cellcolor[HTML]{EFEFEF}RIoTBench    &           & \checkmark             & \checkmark                  &                    & \checkmark           & \checkmark                   & \checkmark                 &                  & \checkmark                   &                               &                      \\ \hline
\cellcolor[HTML]{EFEFEF}SparkBench   & \checkmark         &               &                    &                    & \checkmark           &                     &                   & \checkmark                & \checkmark                   &                               &                      \\ \hline
\cellcolor[HTML]{EFEFEF}CityBench    & \checkmark         & \checkmark             & \checkmark                  & \checkmark                  & \checkmark           &                     & \checkmark                 &                  &                     &                               &                      \\ \hline
\cellcolor[HTML]{EFEFEF}BigBench     & \checkmark         &               &                    & \checkmark                  & \checkmark           &                     &                   & \checkmark                & \checkmark                   &                               &                      \\ \hline
\cellcolor[HTML]{EFEFEF}IoTABench    & \checkmark         &               & \checkmark                  &                    & \checkmark           &                     &                   & \checkmark                & \checkmark                   &                               &                      \\ \hline
\cellcolor[HTML]{EFEFEF}YCSB         & \checkmark         & \checkmark             &                    &                    & \checkmark           &                     &                   &                  & \checkmark                   &                               &                      \\ \hline
\cellcolor[HTML]{EFEFEF}BigDataBench & \checkmark         &               & \checkmark                  &                    & \checkmark           &                     &                   &                  & \checkmark                   &                               &                      \\ \hline
\cellcolor[HTML]{EFEFEF}HiBench      & \checkmark         & \checkmark             &                    &                    & \checkmark           &                     &                   &                  & \checkmark                   &                               &                      \\ \hline
\end{tabular}
\caption{Benchmarks with the different data characteristics they meet. Note that the columns ``data size'' and ``data velocity'' show if the benchmark allows these characteristics to be changed by the user.}
\label{tab:bench}
\end{table}

This list of benchmarks can be used to evaluate different parts of the Smart-* Applications requirements, but none of them provides a common framework. 
In fact, there are only a few benchmarks designed with any of these Smart-* Applications in mind, or at least inspired by them, as seen in Table \ref{tab:benchandusecase}. The benchmarks RIoTBench, IoTABench, LinearRoad and CityBench are some exceptions, covering the Smart Transportation, Smart Grid and Smart Health environments. BigBench and BigDataBench have e-commerce workloads, that cover some of the data processing requirements we explained for Smart Retail, together with datasets that come close to the characteristics expected in Smart-* Applications. However, they do not provide any data on RFID or other types of IoT devices that are present on a Smart Store and focus more on the sales, item stock and customer transactional data. None of the benchmarks actually covers the Smart Factories or the Smart Agriculture use cases. On the other hand, Table \ref{tab:bench} maps the different benchmarks with the general data characteristics and requirements of Smart-* Applications, previously stated in Section \ref{sec:processingrequirements}. This table is intended to be an entry point, from which users can choose the most relevant benchmark for the requirements of their particular use case. However, there are clearly aspects that are not covered by any of the aforementioned benchmarks, such as security and privacy, which are further examined in Section \ref{sec:discussion}. It should be noted that the columns of ``data size'' and ``data velocity'' indicate whether the benchmark allows changing the settings for these parameters (datasets of variable size, changing the data generation rate, etc.).

\begin{table*}[!t]
\scriptsize
\centering
\caption{Benchmarks and the use cases for which their datasets were specifically developed. This table clearly demonstrates the lack of tools to evaluate many of the Smart Environments and Systems}
\label{tab:benchandusecase}

\begin{tabular}{|c|C{1.5cm}|C{1.4cm}|C{1.5cm}|C{1.2cm}|C{1.5cm}|C{1.2cm}|}
\hline
                                              & \cellcolor[HTML]{EFEFEF}\textbf{Smart Resource Usage} & \cellcolor[HTML]{EFEFEF}\textbf{Smart Factories} & \cellcolor[HTML]{EFEFEF}\textbf{Smart Agriculture} & \cellcolor[HTML]{EFEFEF}\textbf{Smart Retail} & \cellcolor[HTML]{EFEFEF}\textbf{Smart  Transportation} & \cellcolor[HTML]{EFEFEF}\textbf{Smart Health} \\ \hline
\cellcolor[HTML]{EFEFEF}\textbf{CloudSuite}   &                                                      &                                                 &                                                   &                                              &                                                       &                                              \\ \hline
\cellcolor[HTML]{EFEFEF}\textbf{LinearRoad}   &                                                      &                                                 &                                                   &                                              & \checkmark                                                    &                                              \\ \hline
\cellcolor[HTML]{EFEFEF}\textbf{RIoTBench}    & \checkmark                                                   &                                                 &                                                   &                                              & \checkmark                                                    & \checkmark                                           \\ \hline
\cellcolor[HTML]{EFEFEF}\textbf{SparkBench}   &                                                      &                                                 &                                                   &                                              &                                                       &                                              \\ \hline
\cellcolor[HTML]{EFEFEF}\textbf{CityBench}    &                                                      &                                                 &                                                   &                                              & \checkmark                                                    & \checkmark                                           \\ \hline
\cellcolor[HTML]{EFEFEF}\textbf{BigBench}     &                                                      &                                                 &                                                   &                 \checkmark                              &                                                       &                                              \\ \hline
\cellcolor[HTML]{EFEFEF}\textbf{IoTABench}     & \checkmark                                                   &                                                 &                                                   &                                              &                                                       &                                              \\ \hline
\cellcolor[HTML]{EFEFEF}\textbf{YCSB}         &                                                      &                                                 &                                                   &                                              &                                                       &                                              \\ \hline
\cellcolor[HTML]{EFEFEF}\textbf{BigDataBench} &                                                      &                                                 &                                                   &  \checkmark                                             &                                                       &                                              \\ \hline
\cellcolor[HTML]{EFEFEF}\textbf{HiBench}      &                                                      &                                                 &                                                   &                                              &                                                       &                                              \\ \hline
\end{tabular}

\end{table*}



\section{Discussion: beyond benchmarks, open issues and future research goals}\label{sec:discussion}
\begin{figure*}
\centering
\includegraphics[scale=0.35]{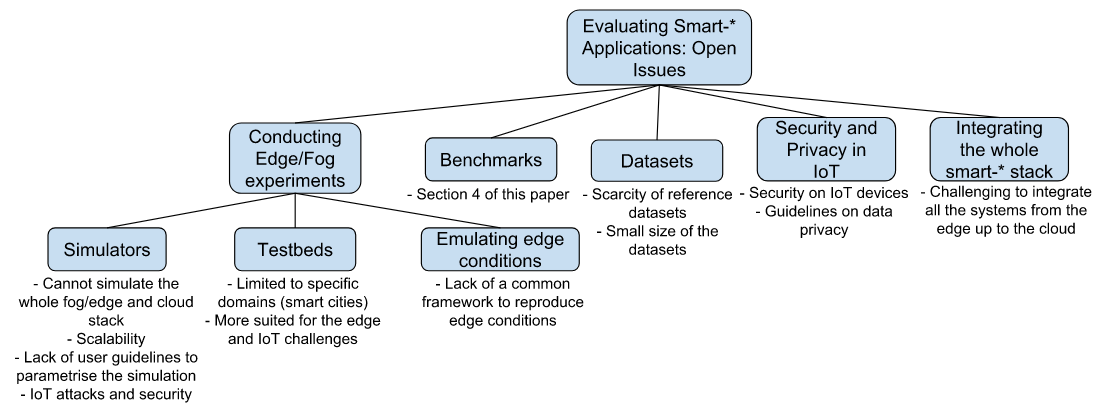}
\caption{The different open challenges and issues when evaluating Smart-* Applications and the systems that support them}
\label{fig:openissues}
\end{figure*}

Smart-* Applications are most probably the next step in using data-powered technologies to improve people's lives, in areas like health, urban planning or transportation among others. To properly evaluate the data processing and storage technologies that form the backbone of these systems, we need a set of tools that ease the simulation of the workloads and the infrastructure supporting the deployment of these applications. In this paper, we have enumerated a set of benchmarks which generate workloads similar to the ones identified for the different Smart-* Applications. One of the main questions is, would this be enough to simulate such complex environments with guarantees? We discuss here a collection of additional tools that can be used to evaluate different Smart-* Applications components, together with a set of open issues, depicted in Figure \ref{fig:openissues}, that need to be addressed in the benchmarking and evaluation aspect to facilitate research in the area. 

\subsection{Conducting Edge and Fog computing experiments}

While a few benchmarks enable researchers to conduct performance evaluations on certain parts of Smart-* Applications, we must underline the scarcity of appropriate testbeds and simulators to conduct experiments under representative conditions, such as Edge computing
Similarly to other areas, we assert that researchers dealing with edge-data analytics challenges should be able to perform their studies on top of experiment driven research frameworks, such as simulators, emulated infrastructures and also in-vivo (i.e. real-world) testbeds.

Simulators allow researchers to get major trends of new algorithms/mechanisms, by implementing straightforward proof-of-concepts. Emulated infrastructures offer control and reproducibility aspects, while relieving researchers from the burden of dealing with the complexity of real infrastructures. Finally, in-vivo testbeds offer a key opportunity for researchers to evaluate proposals (generally the ones that have been validated on top of the two previous infrastructures) under real conditions. This last step is critical to favor the transfer of scientific contributions towards the industry. In the rest of this section, we discuss these types of tools and highlight the most important research efforts in Table \ref{tab:additionaltools}.

\begin{table*}[!t]
\centering
\caption{Research Experiment driven tools to evaluate Smart-* Application Systems}
\label{tab:additionaltools}
\begin{tabular}{|L{4cm}|l|L{5cm}|}
\hline
\rowcolor[HTML]{EFEFEF} 
\textbf{Type of Tool}                                                     & \textbf{Name or authors} & \textbf{Description}                                                                   \\ \hline
\multirow{6}{2cm}[-5.5em]{\textbf{Simulators}} & Rajaram et al. \cite{rajaram2016wireless}          & Health monitoring for smart cities                                                     \\ \cline{2-3} 
                                                                          & S. Karnousk et al. \cite{karnouskos2009simulation}       & Smart Grid with different agents (cars, household, electronic appliances, etc.)        \\ \cline{2-3} 
                                                                          & SCSimulator \cite{santana2016scsimulator}            & Traffic, resource usage and waste management scenarios                                 \\ \cline{2-3} 
                                                                          & CupCarbon  \cite{mehdi2014cupcarbon}              & Deployment of IoT infrastructure (networking, radio modules, transmission power, etc.) \\ \cline{2-3} 
                                                                          & iFogSim   \cite{gupta2017ifogsim}              & Simulates fog hierarchical structure (sensors, gateways, cloud servers, etc.)          \\ \cline{2-3} 
                                    & Silva et al. \cite{silva2013dependability}           & Fault tolerance challenges of IoT                                                      \\ \hline
\multirow{3}{4cm}{\textbf{\small{Emulating a decentralised fog/edge environment}}} & tc$^{1}$                     & Limit bandwidth or introduce latency between nodes                                     \\ \cline{2-3} 
                                                                          & cpulimit$^{2}$, cpupower$^{3}$        & Throttle cpu computing power                                                           \\ \cline{2-3} 
                                         & Enos  $^{4}$                   & Framework to deploy openstack on an emulated fog infrastructure                        \\ \hline
\multirow{3}{2cm}[-0.5em]{\textbf{Testbeds}} & Planet-Lab               & Geo-distributed infrastructure similar to grid computing                               \\ \cline{2-3} 
                                                                          & IoT lab                  & Testbed with sensors, drones and different IoT hardware                                \\ \cline{2-3} 
                                       & Smart Santander          & Smart city testbed                                                                     \\ \hline
\end{tabular}
\end{table*}

\subsubsection{Simulators}
An important step, before the implementation of Smart-* Applications and related services, is to evaluate a prototype or Proof of Concept inside a simulated environment. This allows the users to explore the advantages and the performance of new designs, without incurring the costs of building a physical system, which can be overly large in Smart-* Applications. However, emulating a Smart Environment is not an easy task, especially when we consider that there could be millions of actors (sensors, gateways, servers, users, etc.). Another challenge is that, due to the heterogeneity of this kind of infrastructures, there is an overwhelming amount of information that needs to be included and that is key on evaluating some aspects of Smart-* Applications: energy consumed by the devices, reliability of the connection channels or the different IoT communication protocols, to name a few.     



There have been significant steps in the creation of simulation toolkits for Smart-* Applications. Some of them leverage existing tools, such as Matlab/Simulink\cite{rajaram2016wireless}. Another approach is to develop new systems with that specific purpose on mind. Karnouskos et al.~\cite{karnouskos2009simulation} simulate an smart grid through an agent-based method. SCSimulator\cite{santana2016scsimulator} presents an open-source simulator for smart cities. It focuses on traffic, resource usage and waste management scenarios. CupCarbon\cite{mehdi2014cupcarbon} aims to simulate the deployment of IoT infrastructures. It focuses more on the networking aspect of IoT devices, like broadcasting protocols, using different radio modules in sensors or its transmission power. However, it does not consider the characteristic Fog infrastructure, which is typical of Smart-* Applications. 
This last point is covered by the iFogSim proposal~\cite{gupta2017ifogsim}, which includes many of the elements involved in the typical fog hierarchical infrastructure, such as sensors, actuators, gateways or cloud servers. It is easily extensible and allows the implementation of custom fog applications. However, it leaves out an important assumption in IoT environments, which is that network links are likely to fail or to become unstable. Silva et al.\cite{silva2013dependability} tackle this aspect, presenting a simulator that focuses on the fault tolerance challenges of an IoT infrastructure. 

Throughout all of these publications, the authors acknowledge the need of these kind of simulators in order to evaluate the cost and technological challenges of such a Smart Environment. However, each one focuses on a subset of the requirements we enumerated in Section \ref{sec:characteristics}. In addition, several obstacles can be anticipated from surveying the literature. One of them is the scalability of such systems, specially when they involve millions of actors at the same time. Also, many of these tools allow the user to extend or implement new agents for the simulation, but the community needs guidelines or references about real world scenarios that can lead to a realistic parameterization of those agents (type of sensors, network protocols used, sampling rates, etc.). The security of IoT systems and the different attacks they can suffer is also an important feature that needs to be included. Finally the effects of elasticity on the infrastructure, where devices can leave or join the network, is left out by many of these tools. 

Simulators are extremely important in Smart-* Applications, especially because of the costs involved in deploying such a system in the real world. Our scientific community should increase development efforts on such tools and keep tackling the previously mentioned challenges.

\subsubsection{Emulating a Decentralized Edge/Fog Environment}
Leveraging emulated infrastructures is the second step for validating new algorithms/mechanisms. While applications need to take into account the particularities of this form of processing in their evaluation~\cite{varshney2017demystifying}, the support for specific edge processing requirements is very limited (more often absent) in existing benchmarks, as seen in the previous chapter. This is mainly due to the difficulty of emulating the diverse conditions of edge analytics (e.g. high volatility, churn, limited resources, huge geographical distribution) on today's underlying infrastructures, which mainly rely on the decade old centralized datacenter topology~\cite{nussbaum2017testbeds}. Hence, benchmarking Smart-* Application platforms requires tools and frameworks that can provide the following features:

\begin{enumerate}
\item \emph{Limiting bandwidth between nodes.} Devices at the edge are normally not connected through high speed network interfaces, but through wireless technologies with lower bandwidth. 
\item \emph{Creating geographical areas.} Edge processing is based on the assumption that the communication delay between devices in the same area will be lower. The latency between areas should be included in any Smart Environment benchmark scenario.
\item \emph{Throttling computing power.} Typically, as we move from the edge up to the cloud, the computational capacity of the devices increases.
\item \emph{Mobility.} Some devices in the Edge/Fog can be mobile (e.g. cars or drones). This has implications on their communication patterns and the gateways to which they connect. 
\end{enumerate}

A number of existing dedicated tools can be leveraged to tackle some of these challenges. For instance, \emph{tc}\footnote{\label{foot:tc}\url{http://lartc.org/manpages/tc.txt}} can be used to limit bandwidth or introduce latency between nodes. However, these tools only focus on varying some specific constraints and ignore others (e.g. TCP buffer size, number of outbound requests, etc.). The computing power can be throttled through tools like \emph{cpulimit}\footnote{\label{foot:cpulimit}\url{http://manpages.ubuntu.com/manpages/xenial/man1/cpulimit.1.html}} or \emph{cpupower}\footnote{\label{foot:cpupower}\url{https://linux.die.net/man/1/cpupower}}, the former throttling only by process. This leaves the burden of orchestrating them and applying the most appropriate settings effectively to scientists. Enos\footnote{\label{foot:enos}\url{https://github.com/BeyondTheClouds/enos} (Accessed: March 2018. To be presented to CNERT Workshop, co-located with Infocom 2018)} is a framework that is able to create reproducible Fog/Edge experiments, by specifying a configuration file that deploys the system to be tested on a set of machines. In this configuration file, different network constraints can be specified between machines. However, it only considers OpenStack deployments \cite{sefraoui2012openstack} and it does not include some other features, like computing power throttling.

\subsubsection{\textit{In-vivo} Testbeds}

The absence of specific IoT oriented testbeds has been already cited by other works~\cite{elmangoush2013design}. A real infrastructure with sensors and the gateways attached to them is required to evaluate challenges, like the \emph{security} of the devices or the \emph{heterogeneity} of the data generated by sensors coming from different vendors. Some testbeds that provide a geo-distributed infrastructure exist, such as the Planet-Lab~\cite{chun2003planetlab}. However, this testbed does not provide an IoT environment with sensors, gateways and mobile devices. 
IoT-LAB~\cite{adjih2015fit} provides these features, with a large geo-distributed infrastructure that includes six different sites, 2071 wireless sensors, the possibility of having custom topologies and even drones.
While this testbed provides valuable hardware for experimentation, it lacks the size and complexity of a real world scenario, where the amount of data generated would be much larger. Therefore, it is more suited for Edge and IoT related challenges and does not consider the challenges for Big Data. 
Finally, in the context of Smart Cities use cases, specific testbeds such as SmartSantander~\cite{cheng2015building} have been deployed. SmartSantander provides an urban environment to test and deploy Smart City services. Platforms like these enable the evaluation of prototypes that are to be deployed in real-world environments. However, our scientific community is lacking testbeds that would allow us to investigate more advanced use-cases, composing of several Smart-* Environments/Applications (e.g. Smart Transportation and Smart Health services etc.).

\subsection{Benchmarking security and privacy measures}
Unlike performance and dependability benchmarking, security benchmarking has only been sparsely studied, due to the complexity of designing benchmarks for such non-functional requirements~\cite{Neto2011}. Yet, this is of major importance for Smart-* Applications to become a reality, with roll-outs of smart meters being delayed because of privacy concerns. This problem is exacerbated by the lack of evaluation testbeds that can fully simulate the complex structure of Smart-* Applications and their security vulnerabilities. Another important issue to take into account, when evaluating the data privacy aspect of Smart-* Applications, are the data privacy and protection regulations that exist in many jurisdictions, such as the EU Data Protection Directive~\cite{eu-data} or the Canadian Personal Information Protection and Electronic Documents Act~\cite{pipeda}. Most Smart-* Applications require sensitive data protected under these laws and must take them into consideration when it comes to accessing, processing and transferring this data. As such, evaluating the security and data privacy aspects of Smart-* applications remains an open challenge that requires significant research efforts.

\subsection{Datasets}

Datasets are necessary in order to evaluate any algorithm or new approach to solve a problem. In the field of the Smart-* Applications researchers need more datasets to evaluate new traffic optimisation approaches, health alerting systems or context-aware smart stores to name a few. Although some datasets have been made public \cite{barker2012smart} and open data initiatives \cite{ojo2015tale} encourage efforts on this direction, the datasets are still small in size compared to the overwhelming amount of data that these applications are supposed to generate. In addition, data generation patterns and rates could be provided as complementary information to the dataset, since they are needed to emulate the previously mentioned characteristics of the data in a stream processing scenario (velocity, bursty writes, time windows, etc.).
Finally, datasets can be generated synthetically, provided that the data has the special characteristics that were enumerated in Section \ref{subsec:datacharacteristics}. The previously mentioned IoTABench benchmark can offer a synthetically generated dataset, specifically for the Smart Grid domain. Extrapolating this approach to other Smart-* Domains would facilitate possible experiments in the field.

\subsection{Integrating the whole Smart-* Application stack}

Throughout the current survey, we have shown the different tools available to evaluate the full Smart-* Stack. However, each one covers only specific parts of it. For instance, some benchmarks focus on the data generation part and how it would be ingested by the different Big Data technologies. Some simulators deal with emulating the specific features of sensors and devices at the edge, like the energy consumption or their communication protocols, while others ignore these circumstances and deal only with the data generation part, like the number of data producers, the sampling rate, etc. This complicates the evaluation of the design for the whole application, since we are only able to evaluate the proper functioning either at the edge or at the cloud level. Although this modular approach is beneficial to evaluate each part of the system independently, we cannot ignore the importance of system integration and the process of bringing together all the disparate entities to work as a whole.


\section{Related work}\label{sec:relatedwork}

At the core of Smart-* Applications lies Big Data. As one of the major topics of interests, Big Data has been the focus of an important research effort that is very relevant to the current survey. In particular, a range of great surveys describe the tools commonly used for Big Data processing, as well as the benchmarks to evaluate them, but do not address the specific characteristics of Smart-* Applications.

\subsection{\textbf{Surveys on Big Data benchmarks}}
Being able to evaluate the performance of Big Data applications is critical for evaluating the effectiveness of the data analysis part of a Smart-* stack. The precise challenges associated with this class of applications have been the subject of significant amount of research recently. Yet, most of this work is focused on specific aspects of data processing, and does not seek to provide a complete landscape of benchmarking tools for all aspects of these applications. Shukla \textit{et al.}~\cite{shukla2016benchmarking} provide a precise analysis of the benchmarking tools available for distributed stream processing of IoT applications. A similar work has been performed by Chintapalli \textit{et al.}~\cite{chintapalli2016benchmarking}, detailing the specific aspects of benchmarking stream processing on Spark, Flink and Storm frameworks. Sangroya \textit{et al.} \cite{sangroya2012mrbs} evaluate benchmarking data processing from the angle of dependability. Ivanov \textit{et al.}\cite{ivanov2015big} provide a side-by-side comparison of a variety of Big Data benchmarks for batch processing, providing interesting insights related to their qualities, limitations and characteristics. Designing efficient and comprehensive benchmarks for Smart-* and Big Data applications is difficult; the key requirements, challenges and future directions in developing Big Data benchmarks are detailed by Han \textit{et al.}~\cite{han2014big}~\cite{han2015benchmarking}.

\subsection{\textbf{Surveys on Big Data processing tools}}
Big data has also been studied from the point of view of Smart-* Applications  by Chen et.al.~\cite{chen2014big}, with a paper that provides a complete overview of the different processing models, requirements and frameworks available. A similar work has been done again by the same author, Chen et. al.~\cite{chen2014data}, providing interesting insights of the processing technologies associated with the large volumes of data produced by Smart-* Applications. Bajaber et. al.~\cite{bajaber2016big} draws a generic taxonomy of the processing systems for large-scale Big Data processing. Tsai \textit{et al.} also propose a panorama focused on the tools that one can leverage to extract value from Big Data, both for batch and streaming processing models. Khan~\textit{et al.}~\cite{khan2014big} propose a compelling discussion on the challenges and opportunities that will be faced with Big Data applications. Despite their relevance, these papers only focus on describing the tools associated with Big Data and Smart-* Applications, without describing the tools one can use to benchmark them.

\vspace{1em}

In contrast, we provide a comprehensive overview of the tools one can use to benchmark and evaluate the performance of Smart-* Applications and their associated infrastructure, at each level of the complex processing and storage stack, from the edge to the core.

\section{Conclusions}\label{sec:conclusion}

Driven by the ever-increasing scale of sensor networks, Smart-* Applications are gaining significant importance.  Thanks to rapid advances in underlying technologies, these applications are opening tremendous opportunities for a large number of novel uses, that promise to substantially improve the quality of our lives. While the data volumes they need to process resemble the Big Data movement, these applications have specific requirements of their own. Among these are the characteristics of the data they process, composed of vast amounts of heterogeneous and small data objects, which are flowing at high rate. Furthermore, the promise of near real-time analytics requires to extract knowledge from the data as close as possible to the sensors, leading to extreme geographical distribution of the analytics stack.

Several tools enable users to independently assess the performance of each layer composing a Smart-* infrastructure. In this article, we provide a detailed ensemble review of the tool belt at our disposal. This information is useful to better understand which of the specific requirements of Smart-* Applications is assessed by which tool, enabling users to ensure that a \mbox{Smart-*} Application platform delivers its real-time promise in terms of scale, computational power and responsiveness.

While the tools that are detailed in this article provide a significant help in this difficult task, this survey also highlights the remaining research challenges, in order to encompass all aspects of Smart-* Applications in performance evaluations. Although Fog and Edge computing testbeds and simulators do appear in the literature, a significant amount of work is required to ensure that the whole Smart-* application stack is included in the evaluation process. Moreover, there is a lack of datasets that fully capture the characteristics of data used by Smart-* Applications. Finally, it is critical to precisely evaluate the potential impact of data collected at such a large scale on security and privacy. We strongly believe that answering these open research issues is key to ensuring that, despite their inescapable explosion in scale and complexity, Smart-* Applications continue to deliver the promise of making modern living more enjoyable for everybody.
\section*{Acknowledgments}

All authors contributed equally to this work. This work is part of the "BigStorage: Storage-based Convergence between HPC and Cloud to handle Big Data" project, \textit{H2020-MSCA-ITN-2014-642963}, funded by the European Commission within the Marie Skłodowska-Curie Actions framework. We thank the anonymous reviewers who provided valuable counsel and expertise, that greatly helped this research.

\bibliographystyle{unsrt}  
\bibliography{references} 

\vspace{\baselineskip}
\vspace{\baselineskip}
\vspace{\baselineskip}

\begin{wrapfigure}{l}{25mm} 
    \includegraphics[width=1in,height=0.8in,clip,keepaspectratio]{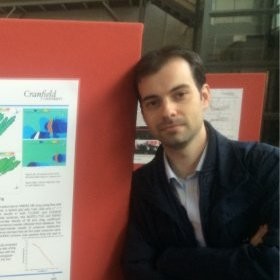}
  \end{wrapfigure}\par
  \vspace{\baselineskip}
  \textbf{Athanasios Kiatipis} received the B.Sc. degree in Mechanical Engineering from the Piraeus University of Applied Sciences, Greece, in 2011, and his Master Degree in Computational Fluid Dynamics from Cranfield University, England, in 2014, where he received an AeroMSc Scholarship from the Royal Academy of Engineering. He is employed Fujitsu Technology Solutions in Munich, Germany, initially as a holder of a "Marie Sklodowska-Curie H2020 ITN Scholarship" and currently as a technology consultant for Artificial Intelligence topics. His current research interests include reinforcement learning, artificial intelligence, computer vision and production manufacturing systems.\par
\vspace{30pt}

  \textbf{} \par
\begin{wrapfigure}{l}{25mm} 
  \includegraphics[width=1.25in,height=0.9in,clip,keepaspectratio]{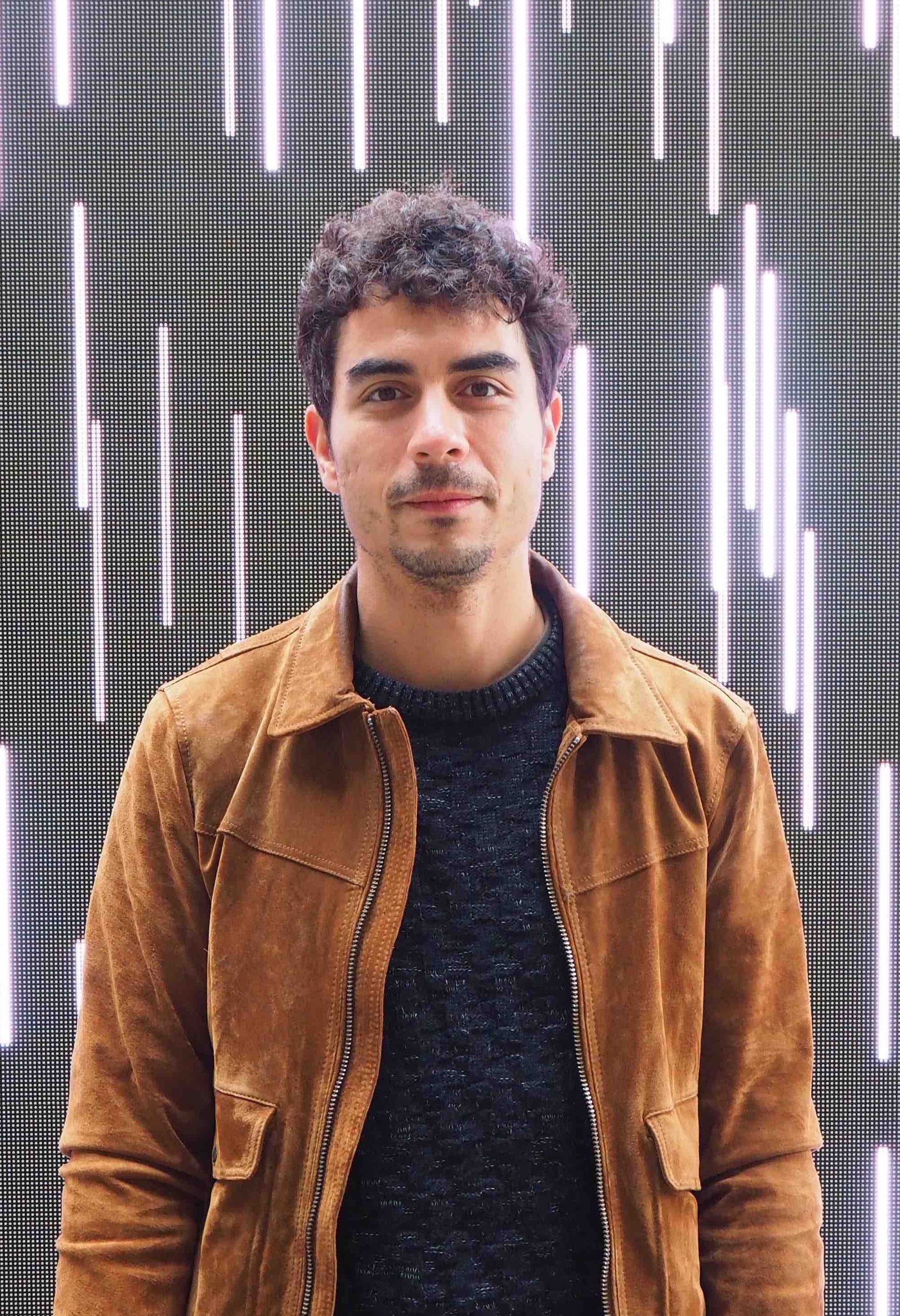}
\end{wrapfigure} \par
\textbf{Álvaro Brandón} is a Ph.D. Student at the Universidad Politécnica de Madrid. He is also a Marie Curie Fellow in the Big Storage H2020 ITN, whose main goal is to train future data scientists in order to take advantage of an overwhelmed data world. He graduated in Computer Science from the University of Leon in 2008. After working in the industry, he completed his Master Degree in Data Science and Analytics at University of Cork, Ireland in 2015. His research is focused on the development of decision support systems to manage complex environments, such as Big Data frameworks like Apache Spark or container orchestrator tools like Kubernetes. His main interests are machine learning, recommendation systems and new distributed computing paradigms, like Fog computing.\par
\vspace{30pt}
  
  \textbf{} \par

\begin{wrapfigure}{l}{25mm} 
    \includegraphics[width=0.7in,height=1.25in,clip,keepaspectratio]{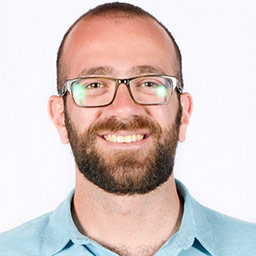}
\end{wrapfigure}\par
  \textbf{Rizkallah Touma} is a Big Data researcher and engineer. He obtained his Ph.D. in Computer Architecture from the Universitat Politècnica de Catalunya (UPC) in 2019 and was a holder of a Marie Sk{\l}odowska-Curie H2020 ITN Scholarship. He received a joint MSc degree in Business Intelligence from the Université Libre de Bruxelles (ULB) and Universitat Politècnica de Catalunya (UPC) in 2015 and a BSc in Computer Science from the University of Damascus, Syria in 2012. His research interests include Big Data Management and Analytics, NoSQL Databases and Semantic Web and Ontologies.\par

\newpage
\vspace{30pt}
\begin{wrapfigure}{l}{25mm} 
    \includegraphics[width=0.7in,height=1.25in,clip,keepaspectratio]{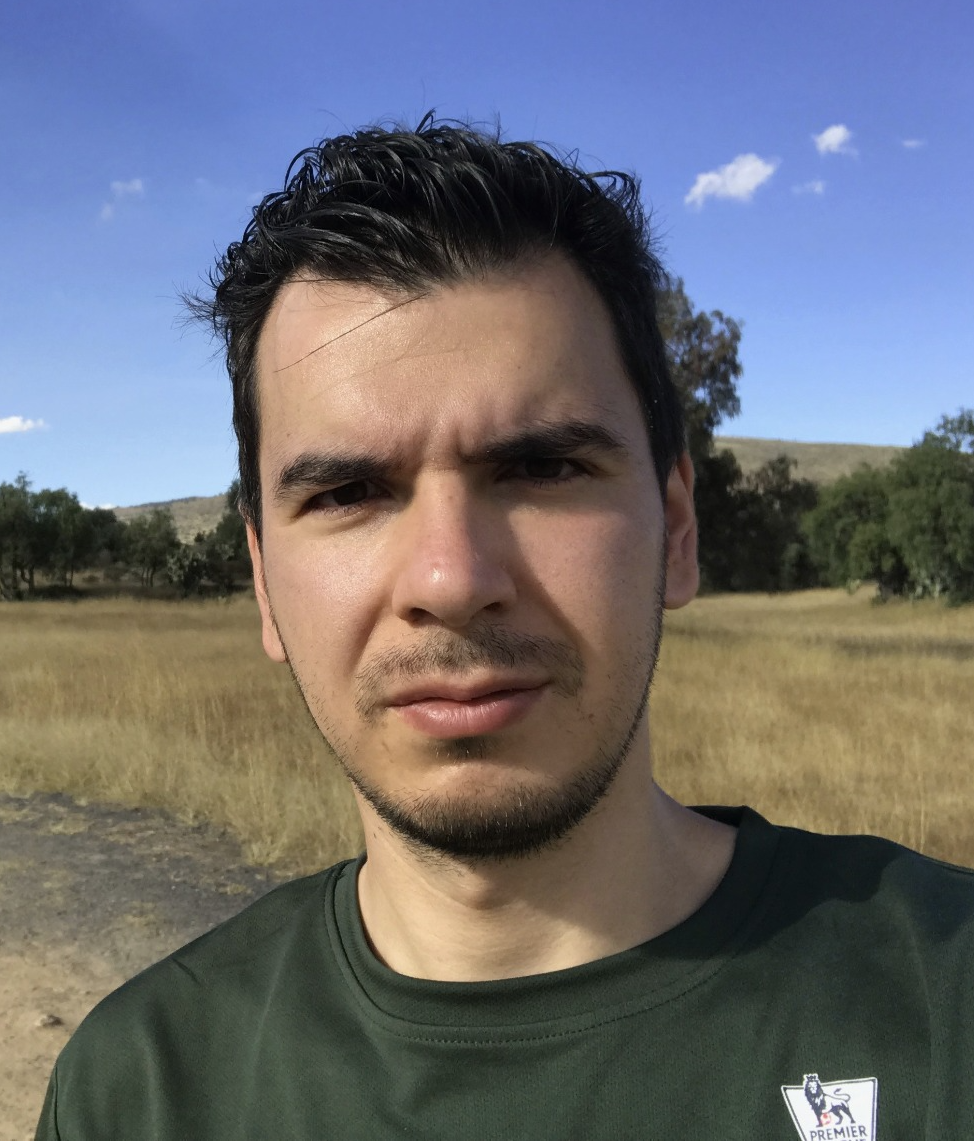}
\end{wrapfigure}\par
  \textbf{Pierre Matri} is a PhD Student at the Ontology Engineering Group since 2015, within the ITN BigStorage H2020 project. He holds a Master Engineering Degree in Software Engineering from the University of Savoie (France, 2009). He previously worked as a Research Engineer at Inria (French Institute for Research in Computer Science and Automation) in Bordeaux and Rennes, in the field of Big Data mining optimisation. He is also working on developing storage systems for billions of small data objects.\par

\vspace{40pt}
  
  \textbf{} \par
  
\begin{wrapfigure}{l}{25mm} 
    \includegraphics[width=0.6in,height=0.7in,clip,keepaspectratio]{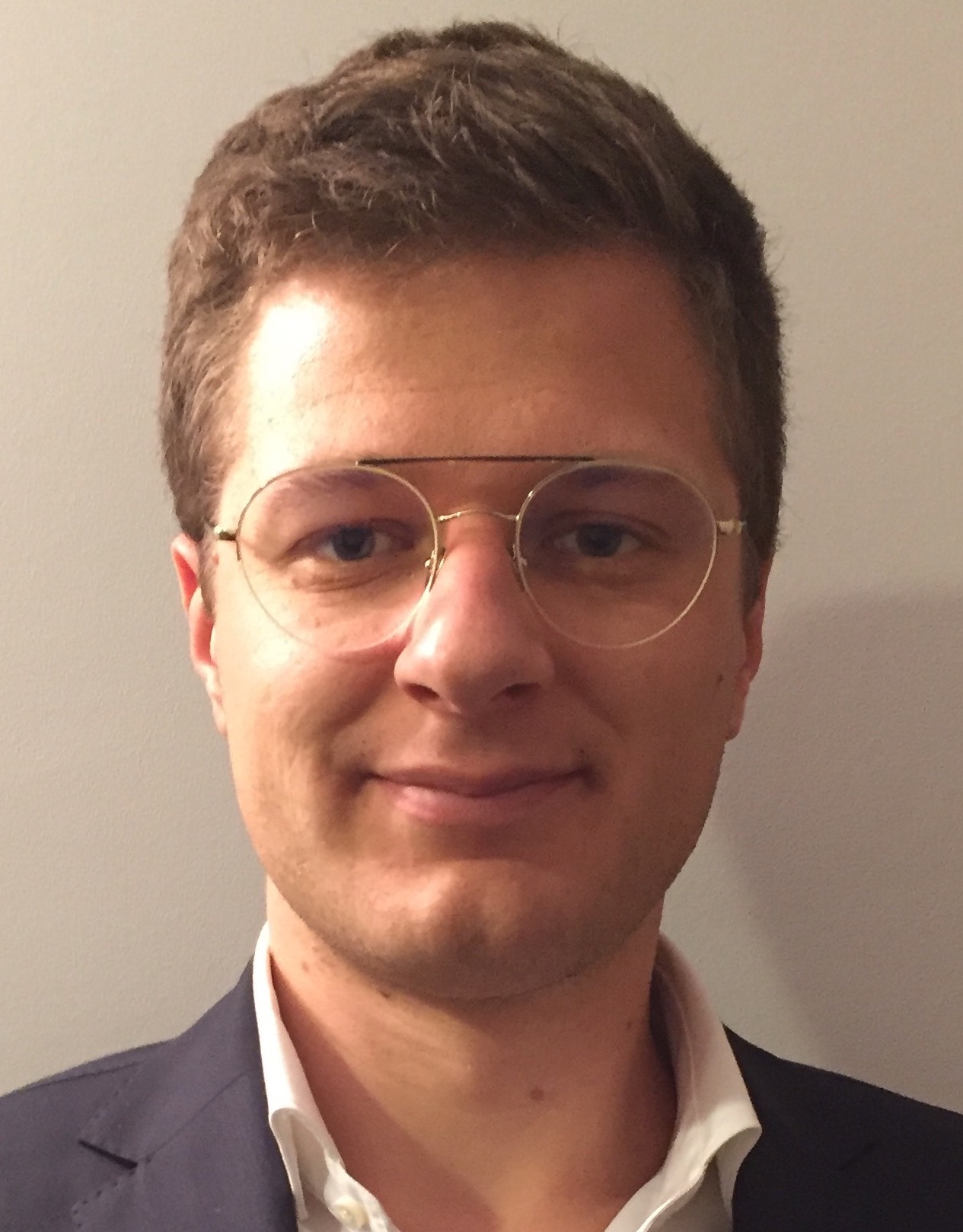}
\end{wrapfigure}\par
  \textbf{Dr. Micha{\l}~Zasadzi{\'n}ski} obtained a Doctorate Degree \textit{cum laude} in Computer Architecture at Universitat Politecnica Catalunya in 2018. He has years of experience in R\&D of business software, especially in Big Data and Machine Learning. His principal interests are automated root cause analysis and intelligent diagnostics of complex distributed systems, including data centers and Big Data systems. His work includes advanced reasoning systems for the distributed computing environments.\par
\vspace{40pt}

\textbf{} \par

\begin{wrapfigure}{l}{25mm} 
    \includegraphics[width=0.7in,height=0.5in,clip,keepaspectratio]{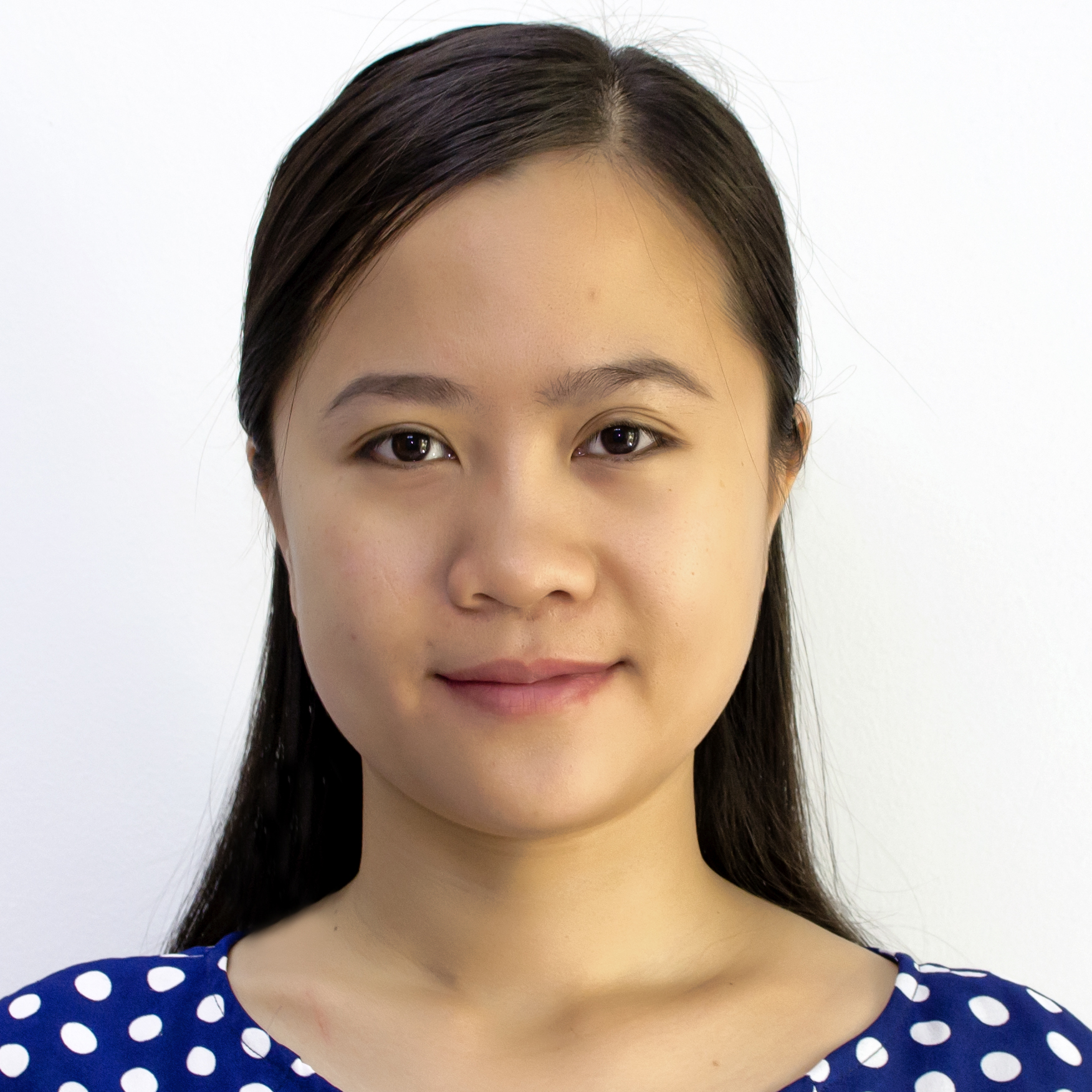}
\end{wrapfigure}\par
  \textbf{Thuy Linh Nguyen} received her Master degree from the University of Science of Ho Chi Minh in Vietnam after doing a 6-month internship at the National Institute of Informatics in Tokyo, Japan. She is currently a PhD student at the INRIA Rennes - Bretagne Atlantique in France, in the European BigStorage H2020 project. Her research focuses on virtualization technologies and storage techniques.\par
\vspace{70pt}
\textbf{} \par
\begin{wrapfigure}{l}{25mm} 
    \includegraphics[width=1in,height=0.7in,clip,keepaspectratio]{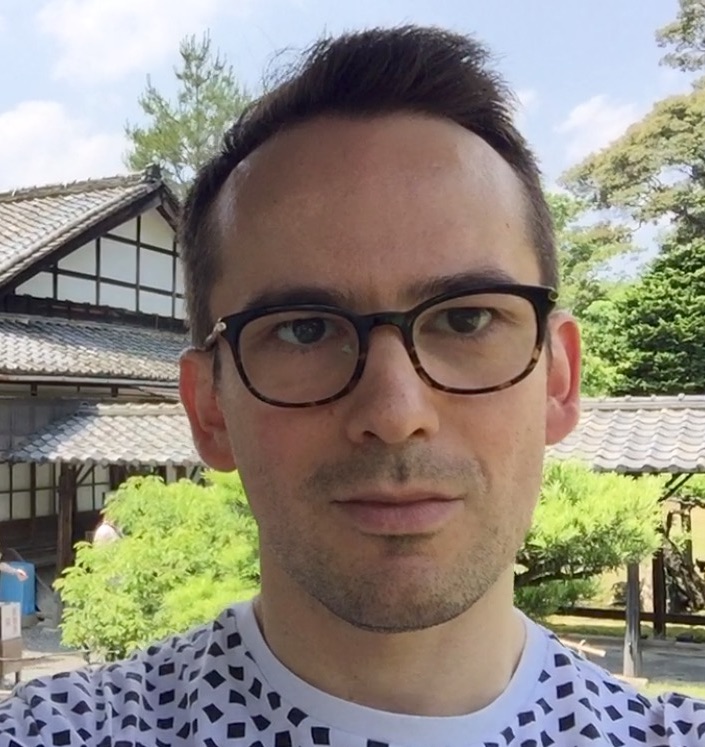}
  \end{wrapfigure}\par
  \textbf{Dr. Alexandru Costan} is an Associate Professor at INSA Rennes and a researcher within the KerData team at Inria/IRISA Rennes, France. Prior to joining Inria, he received his Ph.D. degree in Computer Science in 2011 from the University Politehnica of Bucharest for a thesis focusing on self-adaptive behavior of large-scale distributed systems based on monitoring information, being one of the main creators of the MonALISA monitoring system. His research interests include Big Data management on large scale infrastructures like grids and clouds, fast data and event streaming, autonomic behavior and workflow management.\par
  \vspace{20pt}
\textbf{} \par

\begin{wrapfigure}{l}{25mm} 
    \includegraphics[width=1in,height=0.7in,clip,keepaspectratio]{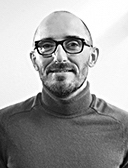}
  \end{wrapfigure}\par
  \textbf{Adrien Lebre} is a full Professor at IMT Atlantique, Nantes (France) and head of the STACK Research Group. He received his PhD from Grenoble Institute of Technologies. His activities aim at designing and implementing new distributed systems in two specific domains in particular: storage systems and virtualisation technologies for cloud architectures and beyond. Since 2015, his activities have been mainly focusing on the Edge Computing paradigm, in particular in the OpenStack ecosystem. \par

\end{document}